\documentclass{aa} 
\usepackage{graphicx}
\usepackage{txfonts}
\begin{document}
   \title{SiO collimated outflows driven by high-mass YSOs in
   G24.78+0.08}


   \titlerunning{SiO collimated outflows in G24.78+0.08} 

   \author{
Codella C. \inst{1,2} \and 
Beltr\'an M.T. \inst{1} \and 
Cesaroni R. \inst{1} \and
Moscadelli L. \inst{1} \and  
Neri R. \inst{3} \and
Vasta M.  \inst{1} \and
Zhang Q. \inst{4}
}

\institute{
INAF, Osservatorio Astrofisico di Arcetri, Largo E. Fermi 5, I-50125 Firenze, Italy
\and
UJF-Grenoble 1 / CNRS-INSU, Institut de Plan\'etologie et d'Astrophysique de Grenoble (IPAG) UMR 5274,
Grenoble, F-38041, France
\and
IRAM, 300 rue de la Piscine, 38406 Saint Martin d'H\`eres, France
\and
Harward-Smithsonian Center for Astrophysics, 60 Garden Street, Cambridge MA 02138, USA
}

   \date{Received date; accepted date}

\abstract 
{The region G24.78+0.08, which is associated with a cluster of high-mass young stellar objects in different 
evolutionary stages, is one of the best laboratories to investigate 
massive star-formation.}
{We aim to image the molecular outflows towards G24.78+0.08
at high-angular resolution using SiO emission, which is considered the classical tracer of protostellar jets.
In this way we study the mass loss process in which we previously detected a 
hypercompact ionised region, as well as rotation and infall signatures.} 
{We performed SiO observations with the VLA interferometer in the $J$ = 1--0 $v$=0 transition and with the
SMA array in the 5--4 transition. A complementary IRAM 30-m single-dish survey in the (2--1), (3--2),
(5--4), and (6--5) SiO lines was also carried out.} 
{Two collimated SiO high-velocity (up to 25 km s$^{-1}$ w.r.t. the systemic
velocity) outflows driven by the A2 and C millimeter continuum  
massive cores have been imaged. On the other hand, we detected no
SiO outflow driven by  
the young stellar objects in more evolved evolutionary
phases that are associated with ultracompact (B) or hypercompact (A1)
H{\sc ii} regions. The A2 outflow has also been traced using H$_2$S. 
The LVG analysis of the SiO emission reveals high-density gas 
(10$^{3}$--10$^{4}$ cm$^{-3}$), with well constrained SiO column
densities (0.5--1 10$^{15}$ cm$^{-2}$).
The driving source of the A2 outflow is associated with typical
hot core tracers such 
as CH$_3$OCHO (methyl formate), C$_2$H$_3$CN (vinyl cyanide),
HCC$^{13}$CN (cyanoacetilene), and (CH$_3$)$_2$CO (acetone).} 
{The driving source of the main SiO outflow in
G24 has an estimated luminosity of a few 10$^{4}$ $L_{\rm \sun}$ (typical of
a late O-type star) and is embedded in the 1.3 mm continuum core A2, which in turn 
is located at the
centre of a hot core that rotates on a plane perpendicular to the outflow main axis.
The present SiO images
support a scenario similar to the low-mass case for massive star formation, where
jets that are clearly traced by SiO emission, create outflows of swept-up ambient gas 
usually traced by CO.} 
 

   \maketitle
%

\def\HII{H{\sc ii}}
\def\WAT{H$_2$O}
\def\MET{CH$_3$OH}
\def\AMM{NH$_3$}
\def\MCN{CH$_3$CN}
\def\kms{km~s$^{-1}\,$}
\def\Tkin{$T_{\rm kin}$}
\def\T54{$T_{\rm B}$(5--4)}
\def\Tb{$T_{\rm B}$}
\def\Nsio{$N_{\rm SiO}$}
\def\NdV{$N_{\rm SiO}/\Delta V$}
\def\cm{cm$^{-3}$}
\def\cmsq{cm$^{-2}$}
\def\kms{km~s$^{-1}$}
\def\H2{H$_2$}
\def\nH2{$n({\rm H_2})$}
\def\nh{$n_{\rm H}$}
\def\Msun{M$_\odot$}
\def\Rsun{R$_\odot$}
\def\Lsun{L$_\odot$}
\def\Msunyr{M$_\odot$~yr$^{-1}$}
\def\Mj{${\dot M}_{\rm jet}$}
\def\e{{\rm e}}
\def\Thii{\mbox{$T_{\rm HII}$}}
\def\Tex{\mbox{$T_{\rm ex}$}}
\def\Tb{\mbox{$T_{\rm B}$}}
\def\Te{\mbox{$T_{\rm e}$}}
\def\Ta{\mbox{$T_{\rm a}$}}
\def\taue{\mbox{$\tau_{\rm e}$}}
\def\taua{\mbox{$\tau_{\rm a}$}}
\def\Os{\mbox{$\Omega_{\rm S}$}}
\def\Ob{\mbox{$\Omega_{\rm B}$}}

\section{Introduction}

Two main theoretical scenarios, based on accretion, are proposed
to explain the formation of O-B type stars:
(i) the core accretion model (McKee \& Tan
2002, 2003), where massive stars form from massive cores, and (ii) the competitive
accretion model (Bonnell et al. 2007), where a molecular
cloud fragments into low-mass cores, which form
stars that compete to accrete mass from a common gas reservoir.
Both models predict the existence of accretion disks
around massive young stellar objects (YSOs), and the
presence of jets driving molecular outflows.
The core accretion model
is a scaled-up scenario of low-mass star formation.
The competitive accretion
model suggests that massive stars always form in
densely clustered environments and that disks and collimated jets
are perturbated by interaction with stellar companions.
Observation of YSOs with disk/jet systems, and of their properties, would
help to distinguish between models. 

The region G24.78+0.08 (hereafter G24), located at 7.7 kpc from the Sun, is one
of the best laboratories to investigate the process of massive star formation.
Several observational campaigns were performed with single-dish antennas
and interferometers (Codella et al. 1997; Furuya et al. 2002; Cesaroni et al. 2003;
Beltr\'an 2004, 2005, 2006, 2007, 2011; Moscadelli et al. 2007; Vig et al. 2008)
toward this region.
G24 is associated with a cluster of high-mass YSOs
in different evolutionary stages, distributed
in a region with size $\sim$ 10$\arcsec$ (see YSOs labelled A1, A2, B, C, and D in
Fig. 1). Beltr\'an et al. (2005, 2011) resolved A1 and A2 into five 
cores (see labels in Fig. 2) aligned in a southeast-northwest direction, 
which suggests a preferential direction for star formation in this region.
The A cores are embedded in huge (0.1 pc), massive (a few 100 M$_{\rm \sun}$) toroids, 
that rotate around the southeast-northwest (SE-NW) direction, and 
could still host elusive accretion disks in their interior (Beltr\'an et al. 2005). 
This possibility would be even more intriguing for the object labelled A1, which 
is unique, because of the
simultaneous presence of almost all the ''ingredients'' expected in a typical star
formation ``recipe'': a 20~$M_\odot$ star surrounded by a
hypercompact \HII\ region, that is located at the centre of a rotating toroid 
undergoing infall towards the star.
The association of the rotation of toroids with the coherent motion of accretion
disks would be supported by the detection of outflows, possibly driven by jets, 
along the SE-NW direction. 
The occurrence of outflows 
towards A and C has been first reported by Furuya et al. (2002) and has recently been imaged 
by Beltr\'an et al. (2011) at high-angular resolution using CO emission.
In particular, Beltr\'an et al. (2011) suggested that A2, and not A1, is 
the driving source of a bipolar CO outflow located along the
SE-NW direction.
However, given the complex outflow structure towards cores A1 and A2,
we cannot entirely discard the possibility that the core A1 could be powering
an additional outflow in the region.

This doubt calls for high-angular resolution observations of
a reliable jet tracer. 
Silicon monoxide (SiO) thermal emission is the best tool for this purpose: 
unlike other species such as CO, it is associated with shocks  
inside jets, suffers minimal contamination from infalling envelopes or swept-up
cavities, and traces regions close to protostars,
which are heavily extincted even in the near- and mid-infrared regions
(e.g. Codella et al. 2007; Lee et al. 2007a).
Only SiO line emission will allow us to probe the mass loss process
and unambiguously determine the direction of the flows driven by the G24 cluster.
The formation of SiO is attributed to
the sputtering of Si atoms from charged grains in a magnetised C-shock
with velocities higher than 20 km s$^{-1}$
(Schilke et al. 1997; Gusdorf et al. 2008ab).
Although high-angular resolution studies of SiO in high-mass star-forming regions
still refer to a quite limited number of objects 
(Hunter et al. 1999; Cesaroni et al. 1999; Qiu et al. 2007; Zhang et al. 2007),
they confirm the power of SiO in tracing the mass loss process in complex
environments like those typical of the massive star-forming regions.

In this paper, we present SiO(1--0) and SiO(5--4) images obtained  with 
the NRAO\footnote{The National Radio Astronomy Observatory is a
facility of the National Science Foundation operated under cooperative agreement by
Associated Universities, Inc} Very Large Array (VLA) and 
SubMillimeter Array (SMA)\footnote{The SubMillimeter Array is a joint
project between the
Smithsonian Astrophysical Observatory and the Academia Sinica Institute of
Astronomy and Astrophysics, and is funded by the Smithsonian Institution and the
Academia Sinica.} as well as  
a complementary IRAM\footnote{IRAM is supported by INSU/CNRS (France), MPG
(Germany), and IGN (Spain).} 30-m observations to unveil the mass loss 
process driven by the G24 cluster of high-mass YSOs.

\section{Observations}

\begin{table*}
\caption[] {List of the continuum and SiO observations}
\label{parobs}
\begin{tabular}{lccccccccc}
\hline
\multicolumn{1}{c}{Observation} &
\multicolumn{1}{c}{Telescope} &
\multicolumn{1}{c}{$\nu$$^{\rm a}$} &
\multicolumn{1}{c}{$E_{\rm u}$$^a$} &
\multicolumn{1}{c}{$S\mu^2$$^a$} &
\multicolumn{1}{c}{HPBW} &
\multicolumn{1}{c}{PA} &
\multicolumn{1}{c}{Spectral resolution} &
\multicolumn{2}{c}{rms noise$^{\rm b}$} \\
\multicolumn{1}{c}{ } &
\multicolumn{1}{c}{ } &
\multicolumn{1}{c}{(GHz)} &
\multicolumn{1}{c}{(K)} &
\multicolumn{1}{c}{(D$^2$)} &
\multicolumn{1}{c}{(arcsec)} &
\multicolumn{1}{c}{(deg)}  &
\multicolumn{1}{c}{(km s$^{-1}$)} &
\multicolumn{1}{c}{(mJy~beam$^{-1}$)} &
\multicolumn{1}{c}{(mK)} \\
\hline
continuum & VLA-C+D &  43.339 & -- & --  & $1.5\times1.1$  & --11 & -- & 0.8 & 309 \\
continuum & SMA & 219.601 & -- & --  & $1.5\times1.4$  &   74 & -- & 6.0 & 72 \\
SiO(1--0) & VLA-D$^c$  &  43.424 &  2 &  9.6 & $2.2\times1.7$$^c$ & --15$^c$ & 0.67 & 5 & 870 \\
SiO(2--1) & IRAM-30m &  86.847 &  6 & 19.3 & 28 & -- & 0.54 & 77 & 16 \\
SiO(3--2) & IRAM-30m & 130.269 & 13 & 28.9 & 19 & -- & 0.36 & 90 & 18 \\
SiO(5--4) & IRAM-30m & 217.105 & 31 & 48.1 & 11 & -- & 0.43 & 173 & 37 \\
SiO(5--4) & SMA      & 217.105 & 31 & 48.1 & $1.7\times1.4$ & 67 & 0.67 & 36 & 415 \\
SiO(6--5) & IRAM-30m & 260.518 & 44 & 57.7 &  9 & -- & 0.36 & 189 & 42 \\
\hline
\end{tabular}

$^a$ Frequencies and spectroscopic parameters of the molecular
transitions have been extracted from the Jet
Propulsion Laboratory molecular database (Pickett et al. \cite{pickett}).
$^b$ For the molecular line observations the 1$\sigma$ noise is given per
channel.
$^c$ The SiO(1--0) emission has been successfully observed only with the VLA--D configuration. \\ 
\end{table*}

\subsection{VLA}

The G24 cluster was observed with 27 antennas of the NRAO
VLA to measure the SiO(1--0) line emission at 43.4 GHz as well as the continuum emission.
The observations were carried out in the Q-band
with the D-configuration on August 10, 2008, and October 10, 2009,
and with the C-configuration on August 14, 2009.
The half power beam width (HPBW) of the antennas is $\sim$ 1$\arcmin$, which is the field-of-view
of the images. The largest structure visible in the
C+D-configuration is $\sim$ 43$\arcsec$.
The phase reference centre of the observations was set to
$\alpha_{\rm 2000}$ = 18$^{\rm h}$ 36$^{\rm m}$
12$\fs$660, $\delta_{\rm 2000}$ = --07$\degr$ 12$\arcmin$ 10$\farcs$15.

The SiO(1--0) line and the continuum were observed simultaneously
with the correlator in 2 IF mode. The continuum emission was collected
with a 200 MHz bandwidth centred at 43.339 GHz.
The SiO(1--0) line was covered by 
simultaneously using two slightly overlapping  6.25 MHz (43 km s$^{-1}$)
bands\footnote{The 12.5 MHz bandwidth, needed to cover
the whole SiO(1--0) profile, was not available during the observations.}
Unfortunately, technical problems did not allow us to properly trace  
the SiO(1--0) emission in the VLA--D configuration.
The observations were performed in fast-switching mode.
Bandpass and phase were calibrated by observing
1832--105, while the flux density scale
was derived by observing 1331+305 (3C286).
All data editing and calibration were carried out using 
the NRAO AIPS\footnote{http://www.aips.nrao.edu/index.shtml} package. 
The line cubes were obtained by subtracting the continuum
from the line. Images were produced using natural weighting:
Details of the synthesised cleaned beam,
spectral resolution, and rms noise of the maps are given in Table 1.

\subsection{SMA}

The target was observed in the SiO(5--4) line at 217.1 GHz and in the
continuum at 1.4 mm with the SMA using two different array 
configurations. Compact- and extended-array observations were taken on May 21 and July 19, 2008, 
respectively. The covered baselines have lengths between $\sim$ 8 and 100 k$\lambda$ (compact)  
and 17 and 160 k$\lambda$ (extended). 
We used two spectral sidebands, both 2 GHz wide, separated by 
10 GHz, covering the 
frequency ranges of 215.4--217.4 and 225.4--227.4 GHz, with a uniform spectral
resolution of about 0.5 km s$^{-1}$.  
The phase reference centre of the observations was set to 
$\alpha_{\rm 2000}$ = 18$^{\rm h}$ 36$^{\rm m}$
12$\fs$565, $\delta_{\rm 2000}$ = --07$\degr$ 12$\arcmin$ 10$\farcs$90.

Absolute flux calibration was derived from observations of
Titan and Uranus. The bandpass of the receiver was calibrated by 
observations of the quasars 3C279 and 3C454.3. Amplitude and phase calibrations 
were achieved by monitoring 1743--038 and 1911--201. 
We estimated the flux-scale uncertainty to be better than 15\%. 
The visibilities were calibrated with 
the IDL superset MIR\footnote{http://cfa-www.harvard.edu/~cqi/mircook.html}. 
Additional imaging and analysis was performed  
with MIRIAD (Sault et al. 1995) and GILDAS\footnote{http://www.iram.fr/IRAMFR/GILDAS} . The continuum 
was constructed in the (u,v)-domain from the line-free channels both in the LSB and USB. 
Continuum maps were created by combining the 
data of both compact and extended configurations with the ROBUST 
parameter of Briggs (1995) set equal to zero, whereas SiO(5--4) line channel maps  
were created using natural weighting after continuum subtraction in the UV plane.  
Angular and spectral resolution and as map sensitivity are given in Table 1. 

\subsection{IRAM 30-m}

Single-dish observations to prepare the interferometric
observational campaign were obtained with the IRAM 30-m telescope  
at Pico Veleta (Granada, Spain). The observations
were carried out on January 27, 2007 pointing the telescope towards the position
used as phase centre for the SMA observations:
$\alpha_{\rm 2000}$ = 18$^{\rm h}$ 36$^{\rm m}$
12$\fs$565, $\delta_{\rm 2000}$ = --07$\degr$ 12$\arcmin$ 10$\farcs$90.
The pointing was checked by
observing nearby planets or continuum sources and was found to be
accurate to within 3--4$\arcsec$.
The observations were made by position-switching in wobbler mode.
As spectrometer,
an autocorrelator split into different parts 
was used to allow simultaneous observations of four
lines: SiO(2--1) at 86.8, (3--2) at 130.3, (5--4) at 217.1, and (6--5) at 260.5
GHz, respectively (see Fig. 3).
The angular (HPBW) and the velocity resolutions provided by the backend, 
and the reached sensitivities are shown in Table 1. 
The integration time (ON+OFF source) was 70 minutes, while 
the main beam efficiency varies from about 0.77 (at 87 GHz) to 0.48 (at 260 GHz).
The spectra were calibrated with the standard chopper wheel method 
(uncertainty $\sim$ 10\%) and are reported
here in units of main-beam temperature ($T_{\rm MB}$).

\section{Results}

\subsection{Continuum emission}

\begin{table}
\caption[] {Position and peak intensity of the continuum cores$^a$}
\label{table_cont}
\begin{tabular}{lcccc}
\hline
\multicolumn{1}{c}{Core} &
\multicolumn{1}{c}{$\alpha({\rm J2000})$} &
\multicolumn{1}{c}{$\delta({\rm J2000})$} &
\multicolumn{1}{c}{$I^{\rm peak}_{\rm 7mm}$} &
\multicolumn{1}{c}{$I^{\rm peak}_{\rm 1.4mm}$} \\ 
\multicolumn{1}{c}{ } &
\multicolumn{1}{c}{(h m s)}&
\multicolumn{1}{c}{($\degr$ $\arcmin$ $\arcsec$)} &
\multicolumn{2}{c}{(mJy beam$^{-1}$)}  \\ 
\hline
G24 A1 & 18 36 12.55 & --07 12 11.05 & 69 & 422 \\
G24 B  & 18 36 12.65 & --07 12 15.02 & 12 & -- \\ 
G24 C  & 18 36 13.11 & --07 12 07.42 & -- & 95  \\
G24 D  & 18 36 12.17 & --07 12 06.15 & -- & 22 \\
\hline
\end{tabular}

   $^a$ The positions are based on the SMA image for all sources but
   B, which is detected only with the VLA. 
\end{table}

\begin{table}
\caption[] {Positions of the SiO emission peaks in the VLA and SMA images}
\label{table_cont}
\begin{tabular}{lcc}
\hline
\multicolumn{1}{c}{Peak} &
\multicolumn{1}{c}{$\alpha({\rm J2000})$} &
\multicolumn{1}{c}{$\delta({\rm J2000})$} \\
\multicolumn{1}{c}{ } &
\multicolumn{1}{c}{(h m s)} &
\multicolumn{1}{c}{($\degr$ $\arcmin$ $\arcsec$)} \\
\hline
\multicolumn{3}{c}{SiO(1--0) -- VLA} \\
\hline
Red & 18 36 12.55 & --07 12 10.54 \\
Blue$^a$ & 18 36 12.52 & --07 12 09.12 \\
\hline
\multicolumn{3}{c}{SiO(5--4) -- SMA} \\
\hline
Red & 18 36 12.56 & --07 12 10.50 \\
Blue & 18 36 12.45 & --07 12 10.30 \\
\hline
\end{tabular}

 $^a$ The SiO(1--0) blue-shifted emission is affected by absorption at low-velocities (see text).
\end{table}

\begin{table}
\caption[] {List of the species and transitions serendipitously detected}
\label{lines}
\begin{tabular}{lccc}
\hline
\multicolumn{1}{c}{Transition} &
\multicolumn{1}{c}{$\nu$$^{\rm a}$} &
\multicolumn{1}{c}{$E_{\rm u}$$^a$} &
\multicolumn{1}{c}{$S\mu^2$$^a$} \\
\multicolumn{1}{c}{ } &
\multicolumn{1}{c}{(GHz)} &
\multicolumn{1}{c}{(K)} &
\multicolumn{1}{c}{(D$^2$)} \\
\hline
C$_2$H$_5$CN(10$_{\rm 1,10}$--9$_{\rm 1,9}$)  & 86.820 & 24 & 0.5 \\
(CH$_3$)$_2$CO(18$_{\rm 4,14}$--17$_{\rm 5,13}$)-AE & 215.881 & 111 & 655.7 \\
$^{13}$CH$_3$OH--E(4$_{\rm 2,2}$--3$_{\rm 1,2}$) & 215.887 & 45 & 3.5 \\
CH$_3$OCHO-E(20$_{\rm 1,20}$--19$_{\rm 1,19}$) & 215.892 & 298 & 53.0 \\
CH$_3$OCHO-A(19$_{\rm 2,18}$--18$_{\rm 1,17}$) & 216.360 & 109 & 7.2 \\
$^{13}$CH$_3$OH--A(10$_{\rm -2,9}$--9$_{\rm -3,6}$) & 216.370 & 162 & 2.7 \\
H$_2$S(2$_{2,0}$--2$_{1,1}$) & 216.710 & 84 & 2.0 \\
C$_2$H$_3$CN(23$_{\rm 2,22}$--22$_{\rm 2,21}$)  & 216.937 & 134 & 996.4 \\
CH$_3$OH-E(5$_{\rm -1,4}$--4$_{\rm -22}$) & 216.946 & 56 & 1.1 \\
CH$_3$OCHO-E(17$_{\rm 3,14}$--16$_{\rm 3,13}$) & 216.959 & 286 & 43.7 \\
CH$_3$OCHO-E(20$_{\rm 1,20}$--19$_{\rm 1,19}$) & 216.965 & 111 & 52.8 \\
CH$_3$OCHO-A(20$_{\rm 1,20}$--19$_{\rm 1,19}$) & 216.966 & 111 & 52.8 \\
$^{13}$CH$_3$OH--A(10$_{\rm 2,8}$--9$_{\rm 3,7}$) & 217.400 & 162 & 2.7 \\
HCC$^{13}$CN(24--23) & 217.420 & 130 & 346.4 \\
C$_2$H$_5$CN(29$_{\rm 5,25}$--28$_{\rm 5,24}$) & 260.536 & 215 & 417.0 \\
\hline
\end{tabular}

$^a$ Frequencies and spectroscopic parameters of the molecular
transitions have been extracted from the Jet
Propulsion Laboratory molecular database (Pickett et al. \cite{pickett}) for all
transitions except those of methanol, which have been extracted from the
Cologne Database for Molecular Spectroscopy (M\"uller et al. \cite{muller}). \\
\end{table}

Figure 1 shows the VLA and SMA maps of the continuum emission at 7 and 1.4 mm, respectively,
towards G24. Table 2 summarises the position and the peak intensity of the
detected cores. Given the lower angular-resolution, the present continuum image at 1.4 mm does not
add information with respect to our previous PdBI and SMA observations at similar 
wavelengths (1.3--1.4 mm; Beltr\'an et al. 2005, 2011), which are
consistent with the new observations.
On the other hand, the present VLA beam is slightly better than that of our previous
observations at 7 mm (2$\farcs$3$\times$1$\farcs$7, Furuya et al. 2002), but the sensitivity
is a factor 2 lower. Again, the present VLA results agree with the  
previously performed analysis.
The source B has been detected at 7 mm and not 
at 1.4 mm is consistent with
its evolved stage, which is associated with an ultra compact (UC) \HII\ region 
(see e.g. Furuya et al. 2002).
But the analysis of the continuum emission is beyond the
scope of the present paper, and will not be further pursued.

\begin{figure*}
\centering
\includegraphics[angle=-90,width=18cm]{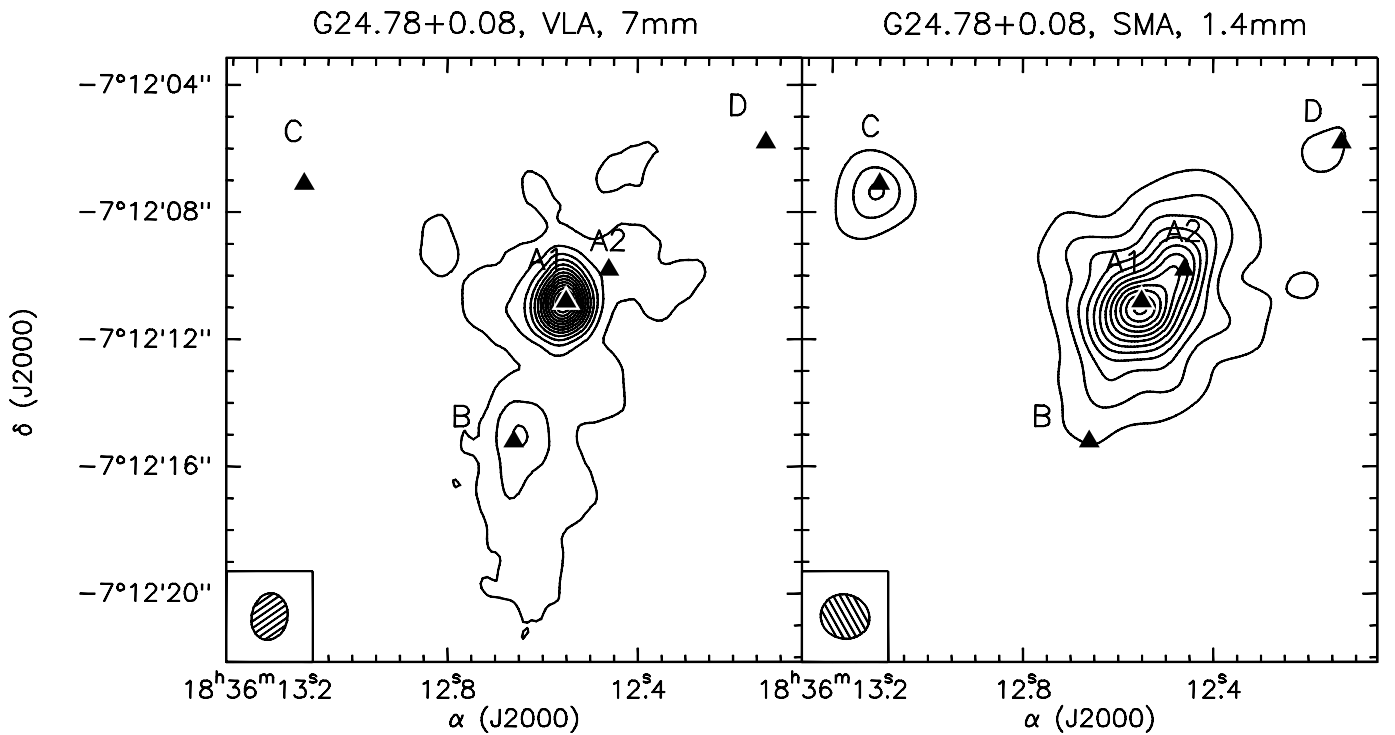}
\caption{Contour plots of the continuum emission at 7 (left panel) and 1.4 mm (right panel) imaged
with the VLA and SMA interferometers, respectively. 
The black triangles mark the positions of the sources identified by Furuya et al. (2002) 
and Beltr\'an et al. (2004).
The 1$\sigma$ rms of the maps is
6.0 (at 1.4 mm) and 0.8 mJy beam $^{-1}$ (at 7 mm), while first contours and
steps correspond to 3$\sigma$ and 6$\sigma$. The ellipses in the bottom-left corners show the HPBW:
$1\farcs5\times1\farcs4$ (PA = 74$\degr$; SMA) and  $1\farcs5\times1\farcs1$ 
(PA = --11$\degr$; VLA).}
\label{mapcontvla}
\end{figure*}

\subsection{SiO outflows}

The outflow activity was previously 
detected using emission of CO isotopologues by Furuya et al. (2002) and Beltr\'an et al. (2011)
with the PdBI and SMA interferometers. Two molecular outflows, both oriented in the
northwest-southest direction, were imaged. One (hereafter called outflow A) associated
with the region of the A1+A2 cores and
is probably driven by A2. 
Another bipolar outflow is associated with core C. 
Figure 2 shows the maps of the integrated blue- and red-shifted SiO(1--0) emission
observed with the VLA in the D configuration towards the G24 cluster, while 
Fig. 4 shows examples of the SiO(1--0) profiles observed towards the positions of A1 and of
the blue- and red-shifted SiO(1--0) emission peaks of the outflow A (see Table 3).
The cloud systemic velocity is +111 km
s$^{-1}$, according to Furuya et al. (2002). The SiO(1--0) lines are characterised by emission up to
high velocities ($\sim$ 20--25 km s$^{-1}$ with respect
to the cloud velocity) in the wings, i.e. velocities  
comparable with what is observed using CO isotopologues. 
For outflow A, the VLA data confirm a poorly collimated 
and extended blue-shifted NW lobe and a smaller and more
collimated red-shifted SE lobe. As seen in CO (Beltr\'an et al. 2011), each lobe is characterised by the
presence of a weaker extended counter-lobe. This could be a geometry effect 
because outflow A's main axis lies close to the plane of the sky.
Alternatively, this could reflect the occurrence of a second outflow.
For outflow C, the SiO(1--0) map shows a very elongated red-shifted
($\sim$ 25 km s$^{-1}$ w.r.t. the systemic velocity) lobe: by dividing the length of the
outflow by its width, we derive a collimation factor $f_c\sim 4$.
In this case, we detect no SiO blue-shifted emission, 
whereas the CO data of Beltr\'an et al. (2011) clearly show 
blue lobe, albeit smaller (by a factor 3) than the red lobe. This suggests 
that outflow C could be located
close to the edge of the molecular cloud and therefore the blue-shifted gas could
flow through a low-density region. This could also explain the lack of SiO emission,
which is expected to trace the high-density collimated wind from the YSO. 

Finally, no outflow
activity has ever been observed towards source B, in agreement with its evolved stage,
which is associated with an UC\HII\ region 
(Codella et al. 1997; Furuya et al. 2002). No SiO emission (or CO emission) 
has been detected towards core D, which remains the most enigmatic object, bevause it is
traced only by $\sim$1--3~mm continuum emission and is not detected in any molecule
(Furuya et al. 2002; Beltr\'an et al. 2011). Our non-detection seems to confirm
that D is a non-centrally-peaked core without active star formation. 

\begin{figure*}
\centering
\includegraphics[angle=-90,width=16cm]{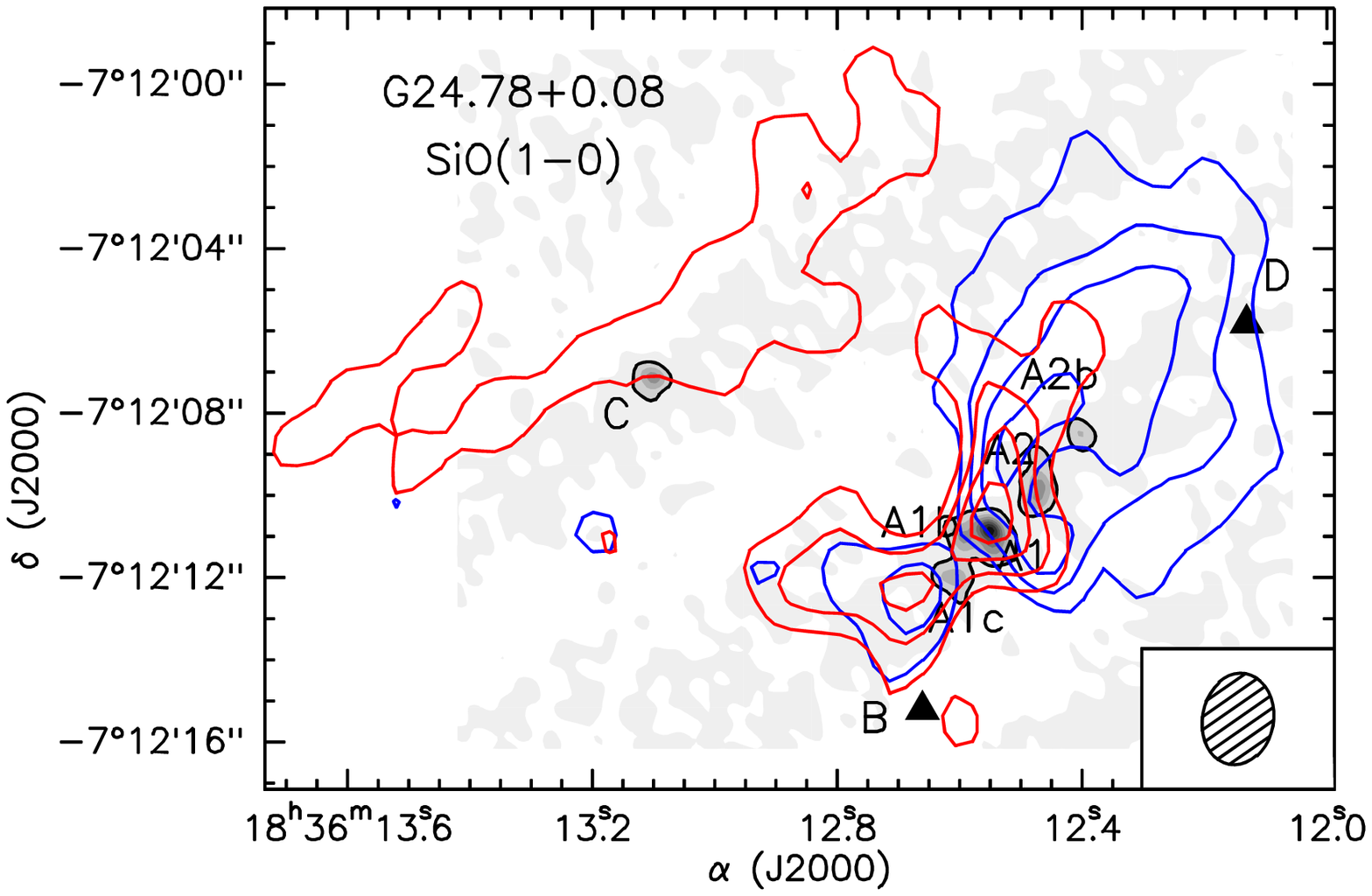}
\caption{Contour map of blue- and red-shifted SiO(1--0) VLA emission 
superimposed on the 1.3 mm continuum emission as observed at high angular 
resolution by Beltr\'an et al. (2011) using a very extended SMA configuration
($0\farcs55\times0\farcs44$). The sources of the G24.78+0.08 cluster  
are labelled following Furuya et al. (2002) and Beltr\'an et al. (2004, 2011). 
The SiO emission was averaged over the velocity intervals (+80,+111) km s$^{-1}$
and (+111,+132) km s$^{-1}$ for the blue- and red-shifted emission, respectively.
The 1$\sigma$ rms of the SiO maps is 1.2 mJy beam$^{-1}$ km s$^{-1}$. 
Contour levels range from 5$\sigma$ by steps of 3$\sigma$.
The filled ellipse in the bottom-right corner shows the
synthesised beam (HPBW): $2\farcs2\times1\farcs7$ (PA = --15$\degr$).}
\label{mapsio10}
\end{figure*}

\begin{figure}
\centering
\includegraphics[angle=-90,width=8cm]{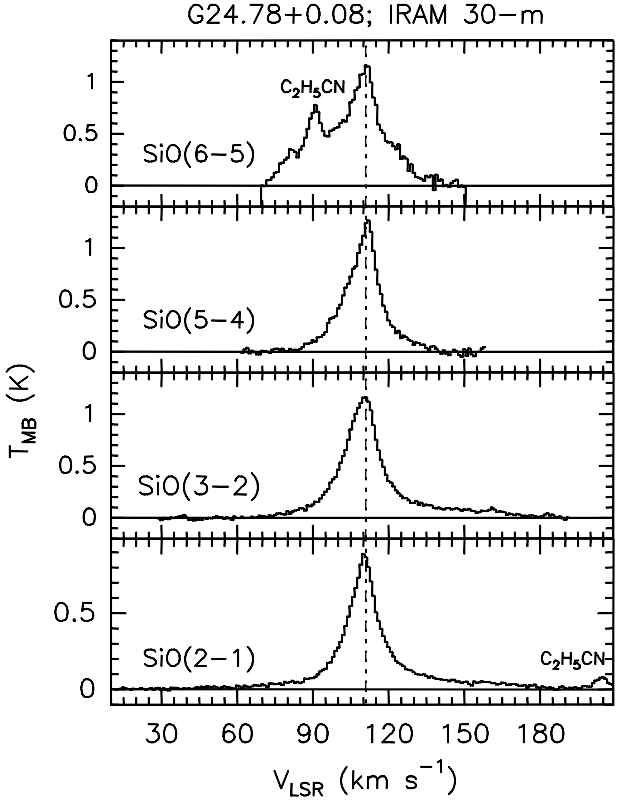}
\caption{SiO line profiles (in main-beam temperature, $T_{\rm MB}$, scale) observed
with the IRAM 30-m antenna towards G24.78+0.08.
The dashed lines stand for the systemic velocity (+111.0 km s$^{-1}$).
Two ethyl cyanide (C$_2$H$_5$CN) emission lines have been serendipitously detected
at 86819.85 MHz (10$_{\rm 1,10}$--9$_{\rm 1,9}$; $E_{\rm u}$=24 K; bottom panel), and
at 260535.69 MHz (29$_{\rm 5,25}$--28$_{\rm 5,24}$;
$E_{\rm u}$=215 K; upper panel, blended
with the SiO(6--5) blue wing).}
\label{spectra30m}
\end{figure}

\begin{figure*}
\centering
\includegraphics[angle=-90,width=18cm]{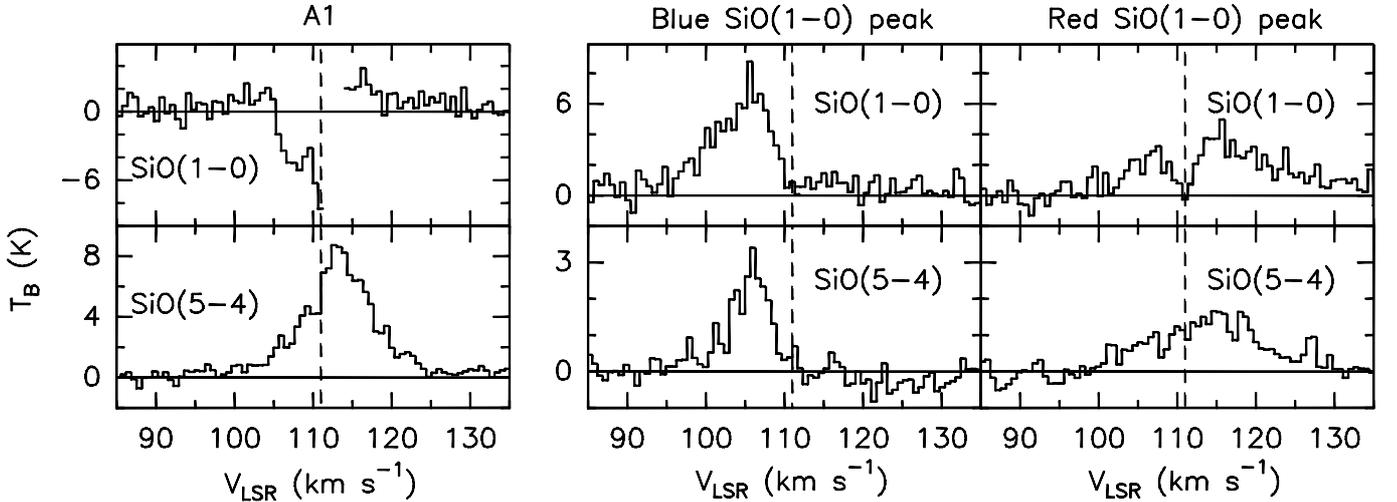}
\caption{Examples of SiO(1--0) and (5--4) line profiles (in brightness temperature
$T_{\rm B}$, scale) observed towards 
A1 (left panel) and the peak positions of the blue- and red-shifted
(middle and right panel) SiO(1--0) line emission. Dashed lines mark the
systemic velocity (+111 km s$^{-1}$).}
\label{spectravla}
\end{figure*}

\begin{figure}
\centering
\includegraphics[angle=-90,width=8cm]{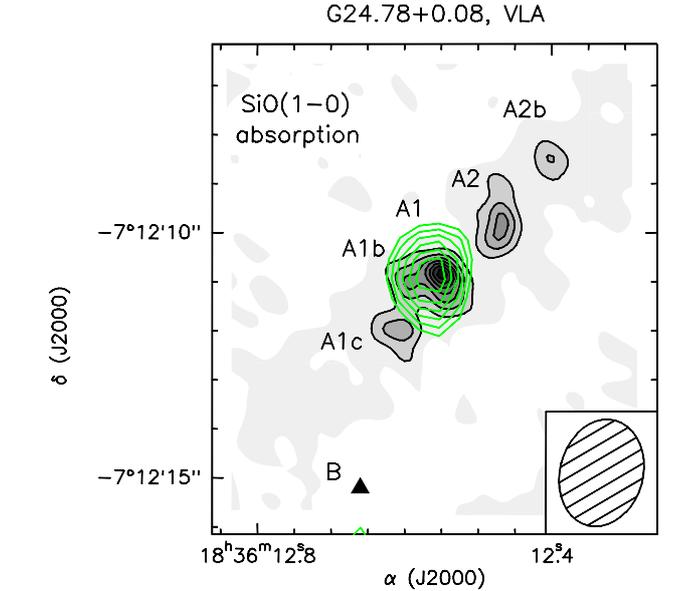}
\caption{Zoom-in of the central region of G24.78+0.08, where
the cluster of A1-A2 continuum sources (grey scale) has been mapped at high angular
resolution by Beltr\'an et al. (2011) using a very extended SMA configuration
($0\farcs55\times0\farcs44$).
The spatial distribution of the
blue-shifted (from +106 to +110 km s$^{-1}$) absorption
(see Fig. 4) observed towards A1 is reported by the green contours.
Contour levels range from --10$\sigma$ to --3$\sigma$ by steps
of 1$\sigma$ (0.3  mJy beam$^{-1}$ km s$^{-1}$).
The filled ellipse in the bottom-right corner shows the
synthesised beam (HPBW): $2\farcs2\times1\farcs7$ (PA = --15$\degr$).}
\label{mapsio10zoom}
\end{figure}

The VLA image of the low-excitation ($E_{\rm u}$ = 2 K) SiO(1--0) emission is affected by absorption 
features\footnote{Unfortunately, 
we lack the information on the
SiO velocity between +111 and +113 km s$^{-1}$ towards the position of A1
because of technical problems.}, which  
prevents us from drawing a definite picture of the morphology of the SiO outflow
and from identifying the driving source.
Figure 5 shows a zoom-in of the central region of G24.78+0.08, centred on  
the cluster of the A1-A2 continuum sources. 
The green contours correspond to the narrow absorption blue-shifted feature 
(from +106 to +110 km s$^{-1}$, see Fig. 4) observed towards A1, 
against the hypercompact \HII\ region
imaged by  Beltr\'an et al. (2007). This is ionised by a zero-age main-sequence star of 
spectral type O9.5 and mass of 
about 20 $M_{\rm \sun}$. The hypercompact \HII\ region is very bright 
and therefore it is expected to observe colder molecular gas in absorption against it.  
Indeed, at cm-wavelengths, where the brightness temperature of A1
is $\sim$ 5 10$^3$ K in a $\sim$ 0$\farcs$8 beam (Beltr\'an et al. 2007), 
the NH$_3$(2,2) spectral pattern has allowed us
to trace the blue-shifted outflow lobe and the red-shifted gas  
infalling towards A1 in absorption (Beltr\'an et al. 2006). 
In conclusion, the VLA SiO(1--0) map calls for the spectral analysis of a
higher-J SiO at higher frequencies, than  
those observed at SMA, where the brightness of the A1 hypercompact \HII\ can be neglected.

\begin{figure*}
\centering
\includegraphics[angle=-90,width=18cm]{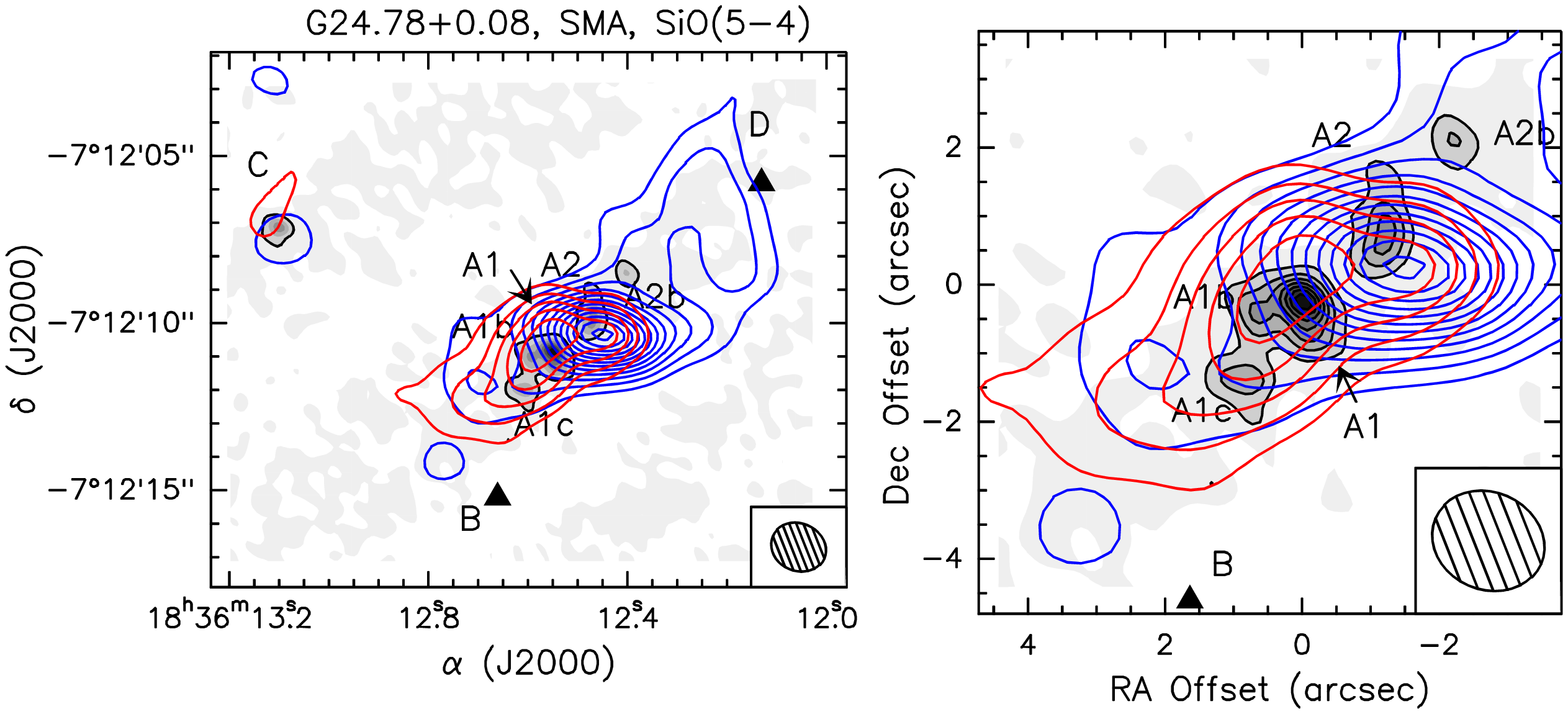}
\caption{Contour map of blue- and red-shifted SiO(5--4) SMA emission
superimposed on the 1.3 mm continuum emission as observed at high angular
resolution by Beltr\'an et al. (2011) using a very extended SMA configuration
($0\farcs55\times0\farcs44$). The Right panel reports a zoom-in of the central
region. The sources of the G24.78+0.08 cluster
are labeled following Furuya et al. (2002) and Beltr\'an et al. (2004, 2011).
The SiO emission was averaged over the velocity intervals (+89,+111) km s$^{-1}$
and (+111,+130) km s$^{-1}$ for the blue- and red-shifted emission, respectively.
The rms 1$\sigma$ of the SiO maps is 18 mJy beam$^{-1}$ km s$^{-1}$.
Contour levels range from 5$\sigma$ by steps of 3$\sigma$.
The filled ellipse in the bottom-right corner shows the
synthesised beam (HPBW): $1\farcs7\times1\farcs4$ (PA =
67$\degr$).}
\label{mapsio54}
\end{figure*}

\begin{figure*}
\centering
\includegraphics[angle=-90,width=16cm]{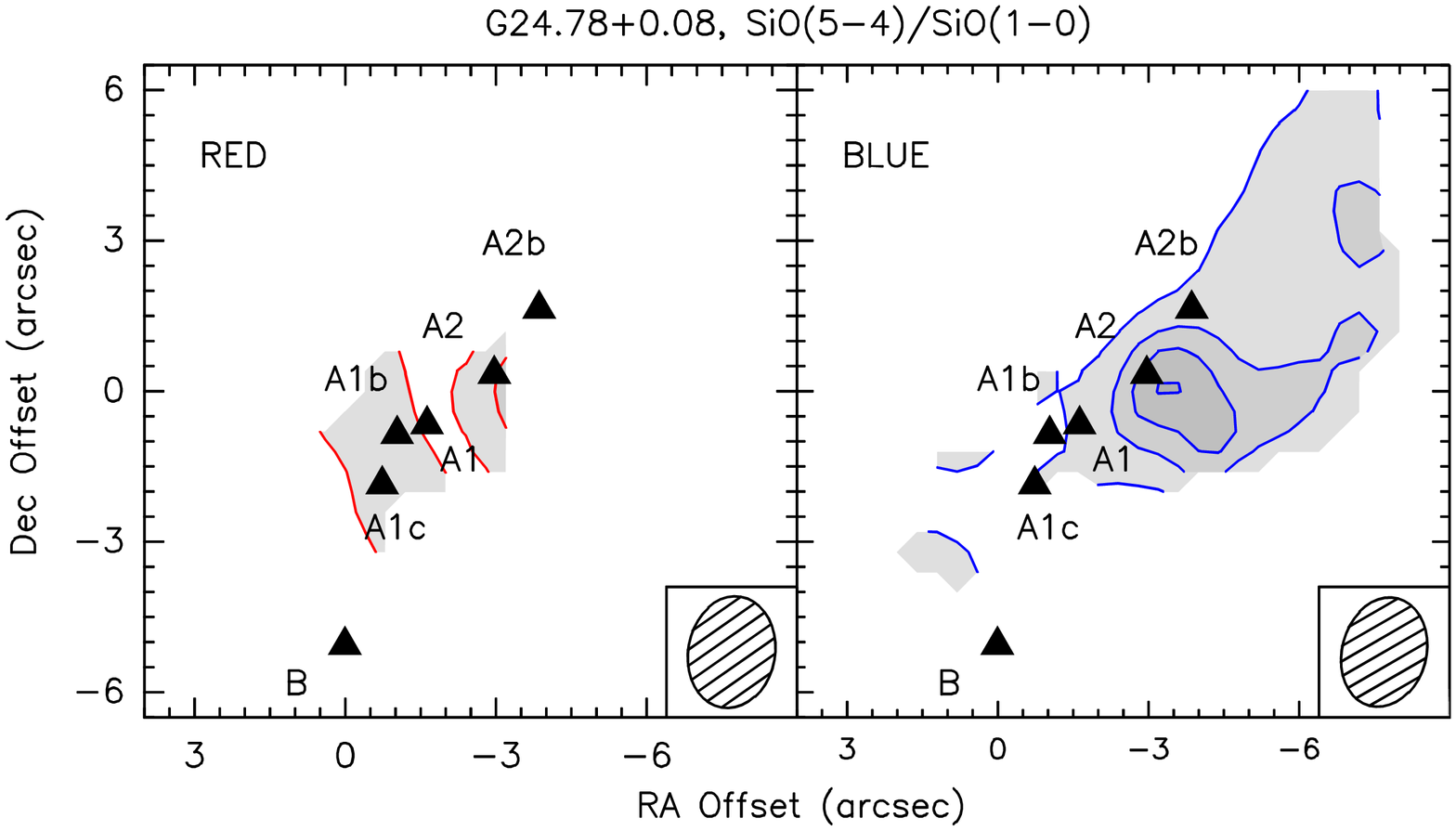}
\caption{SiO(5--4)/SiO(1--0) intensity ratio (brightness temperature, $T_{\rm B}$, scale), 
derived where both emissions have an S/N $\ge$ 3. For this comparison, the SMA image has been smoothed 
to the angular resolution of the VLA image. The velocities affected by SiO(1--0)
absorption (see Fig. 4) have not been taken into account: 
right and left panels are for the blue
(+89,+105 km s$^{-1}$) and red (+115,+130 km s$^{-1}$) emission, respectively. 
Contour levels range from 0.5 to 2.0 by steps of 0.5.
The sources of the G24.78+0.08 cluster
are labelled following Furuya et al. (2002) and Beltr\'an et al. (2004, 2011).}
\label{ratio}
\end{figure*}

Figure 6 shows the SMA maps of the integrated blue- and red-shifted SiO(5--4) emission.
Only a hint of the outflow C is seen,
and, as in the case of SiO(1--0), no emission has been observed towards cores B and D.
On the other hand, the bipolar outflow A is clearly imaged,
with the emission peaks lying closer to the A1+A2 system. 
The SiO(5--4) structure is consistent with that traced by the SiO(1--0) line. 
Table 3 reports 
the positions of the SiO emission peaks detected in the VLA and SMA images.
The position of the red-shifted peak is basically the same (within the uncertainties) 
for the
for the SiO(1--0) and (5--4) lines, whereas, given the 
absorption observed at low velocities that affects the SiO(1--0) map (see Fig. 4), the 
positions of the blue-shifted peaks in the two lines differ by $\sim$ 1$\farcs$6.
Example of SiO(5--4) spectra are shown in Fig. 4.

Figure 7 shows the maps of the SiO(5--4)/SiO(1--0) intensity ratio, obtained after
smoothing the maps to the same angular resolution, and derived where both emissions
have a signal-to-noise (S/N) $\ge$ 3. 
The velocities where SiO(1--0) shows absorption features
(see Fig. 4) have not been taken into account. The red-shifted emission does not  
show a clear trend, while the SiO(5--4)/SiO(1--0) ratio for the blue-shifted emission
peaks in correspondence of A2, suggesting an increase of the
SiO excitation conditions. 
Finally, in Fig. 8 we compare the SiO(5--4) profile observed with the IRAM 30-m antenna, 
with that obtained by integrating the SMA image over the IRAM beam (11$\arcsec$;
blue histogram). The spectral shapes well match and, given
the flux-scale uncertainties of the SMA (15\%) and IRAM 30-m (10\%), the
line intensities also agree well, indicating that the SMA array recovers
at least the 90\% of the emission detected with the single-dish.

\begin{figure}
\centering
\includegraphics[angle=-90,width=8cm]{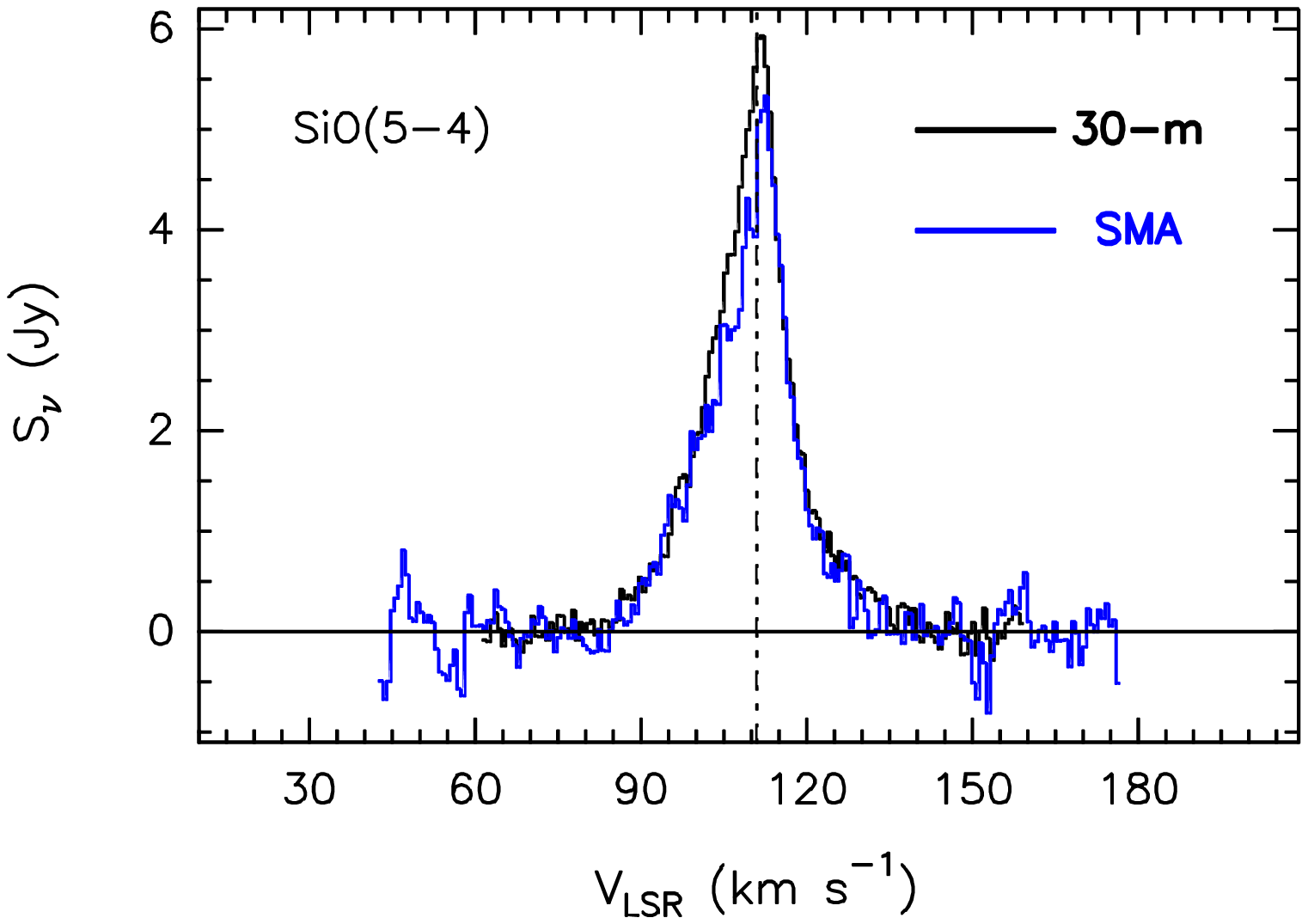}
\caption{Comparison between the SiO(5--4) line profile (in flux density, $S_{\rm \nu}$, scale) observed
with the IRAM 30-m antenna (black) and the spectrum 
obtained from the SMA data (blue) 
by integrating over a region equal to the IRAM 30-m beam (11$\arcsec$). For this 
comparison the IRAM 30-m spectrum has been resampled to the spectral
resolution of the SMA data (0.67 km s$^{-1}$).
The dashed lines stand for the systemic velocity (+111.0 km s$^{-1}$).}
\label{spectra30mcompa}
\end{figure}

\subsection{The origin of the outflow A}

Which is the driving source of the outflow A?
The VLA and SMA SiO images have been overlaid on the 1.3 mm continuum map obtained  
at sub-arcsecond angular resolution ($0\farcs55\times0\farcs44$) by Beltr\'an et al. (2011),
which gives the 
best picture so far available of the YSO population in the G24.78+0.08 cluster. 
It is possible to see that cores A1 and A2 are 
resolved into three (called A1, A1b, A1c) and two (A2, A2b) cores, respectively. 
By analysing the CO emission, Beltr\'an et al. (2011) suggested that the outflow A is actually
powered by core A2, which is massive (22 $M_{\rm \sun}$) and associated with a hot-core
($\sim$ 180 K) detected in CH$_3$CN, a typical hot-core tracer.
The SiO images suggest that A1 is not the driving source of the
outflow motions detected so far. Given the presence of several
YSOs in the region, we cannot exclude additional fainter
outflows. However, the peaks of the SiO(5--4) emission
suggest that the geometrical centre of outflow A, and thus its driving source, 
is located in the southern portion of the A2 core, which
indeed is clearly elongated and probably hosts multiple YSOs.
Additonal evidence comes from the position-velocity (PV) diagram of the SiO(5--4) transition
along the NW--SE axis (see Fig. 9), which seems to confirm that A1 cannot be
the driving source. Interestingly, the average velocity of the SiO emission increases
as a function of distance from the geometrical centre, suggesting gas acceleration.
Figure 10 compares the PV diagram of SiO(5--4) with that derived 
from the CO(2--1) SMA data by Beltr\'an et al. (2011). We can clearly see that
the CO emission (i) is completely absorbed by foreground material at low velocities
and is affected by filtered extended emission, confirming
the need of a tracer such as SiO to properly image the molecular outflow, and (ii) 
shows additional components with respect to what was traced by SiO. Indeed, 
bright CO emission is seen at (--5$\arcsec$,+115 km s$^{-1}$), not traced by SiO, as well as weak
emission at (+2$\arcsec$,+113 km s$^{-1}$). 
In other words, the CO emission cannot be simply
interpreted as an intensity-scaled version of the SiO emission.
These findings are not surprising given what already found for the jets driven
by Sun-like protostars, where, thanks to the smaller spatial scales that can be investigated,
it is possible to see that CO traces not only the jet but also the walls of the
cavity opened by the jet itself (e.g. Lee et al. 2007b, see their Fig. 5). 
In summary, the present SiO maps
support a formation mechanism for massive stars similar to that of their
low-mass counterparts, where
jets, clearly traced by SiO emission, create outflows of swept-up ambient gas 
traced by CO. The possibility of having an SiO jet is also 
supported by the extremely high velocity 
(up to $\sim$ 70 km s$^{-1}$ with respect to the systemic velocity) of the SiO emission
detected in the $J$ = 2--1 and 3--2 spectra thanks to the high sensitivities provided
at these frequencies by the 30-m antenna. 
Of course, in the present case, the angular resolution is not high
enough to assess the occurrence of SiO jets. However, in analogy to the
SiO jets in low-mass stars, one cannot rule out the possibility that
the elongated SiO outflow does trace jet activity.

\subsection{Other molecular species: the outflow and the hot core}

The 2-GHz-wide LSB bandwidth used to trace SiO(5--4) emission with the SMA allowed us
to serendipitously observe several lines of different molecular species,
which are listed in Table 4 and shown in Fig. 11. In particular, the LSB spectrum is dominated  
by broad H$_2$S(2$_{2,0}$--2$_{1,1}$) emission, which confirms that
hydrogen sulphide is an excellent tracer of molecular
outflows (Codella et al. 2003, Gibb et al. 2004). Figure 12 shows the 
integrated blue- and red-shifted H$_2$S emission from outflow A, 
while Fig. 9 shows the corresponding PV diagrams.

Outflow A is clearly seen in the H$_2$S image, which suggests (like SiO) that the
geometrical centre (and thus the driving source) is located
in the southern portion of the A2 core.  
Interestingly, the H$_2$S outflow is definitely less collimated than
the SiO(5--4) one, observed with the same angular resolution.
H$_2$S and SiO show different behaviours also in the PV diagrams, where hydrogen sulphide
does not show a clear increase of the average velocity  
with the distance from the driving source. 
These findings could be the signature of an additional fainter outflow 
driven by one of the A1+A2 YSOs that contributes to the observed emission.
However, what we found agrees with an enhancement of the H$_2$S abundance as a consequence
of the evaporation of the dust mantles in a shocked gas (e.g. van Dishoeck \& Blake 1998), 
whereas SiO comes from a definitely
smaller region that is directly associated with the primary jet, where the
refractory dust cores are disrupted as well. 

In addition to SiO and H$_2$S, we detected several high-excitation
($E_{\rm u}$ between 45 and 162 K) transitions of methanol isotopologues 
(see Table 4). Figure 11 shows the spectra observed
towards the A2 position. The CH$_3$OH line profiles are narrower than those  
of SiO and H$_2$S and, although a weak component related to outflows cannot
be excluded for $^{12}$CH$_3$OH, it appears that methanol emission is 
dominated here by gas heated by
YSOs. Indeed, from Fig. 9 one clearly sees that the PV diagram of methanol,
outlining a compact circular pattern, is
significantly different from that of SiO. Most likely,
the CH$_3$OH molecules trace the hot core A2 (already imaged by Beltr\'an et al.(2005), 
and are released from dust mantles because of stellar irradiation. 
In particular, $^{13}$CH$_3$OH peaks at a position that matches the geometrical
centre of the outflow well (see the solid vertical line in Fig. 9). 
Thus, the present observations suggest that a YSO lying in the SE of core A2
is heating the gas and driving the outflow A. 

Finally, we also detected several lines in the SMA spectral window that are due 
to cyanoacetylene (in HCC$^{13}$CN form) as well as to other complex molecular 
species usually considered as hot-core tracers: CH$_3$OCHO (methyl formate),
C$_2$H$_3$CN (vinyl cyanide), and, tentatively, one transition of (CH$_3$)$_2$CO (acetone). 
Other complex species were reported by  
Beltr\'an et al. (2011, see their Table 1). 
Indeed, these emissions are
characterised by high excitation ($E_{\rm u}$ in the 109--298 K range) and
compact spatial distribution (see the C$_2$H$_3$CN in Fig. 9 as an example),
indicating an association with the hot core traced by $^{13}$CH$_3$OH.
Remarkable is the tentative detection of acetone, given that only recently this organic specie has been
detected towards the hot molecular core Sagittarius B2(N-LMH) and in Orion BN/KL
(Snyder et al. 2002, Friedel et al. 2005, Goddi et al. 2009). 

\begin{figure*}
\centering
\includegraphics[angle=-90,width=18cm]{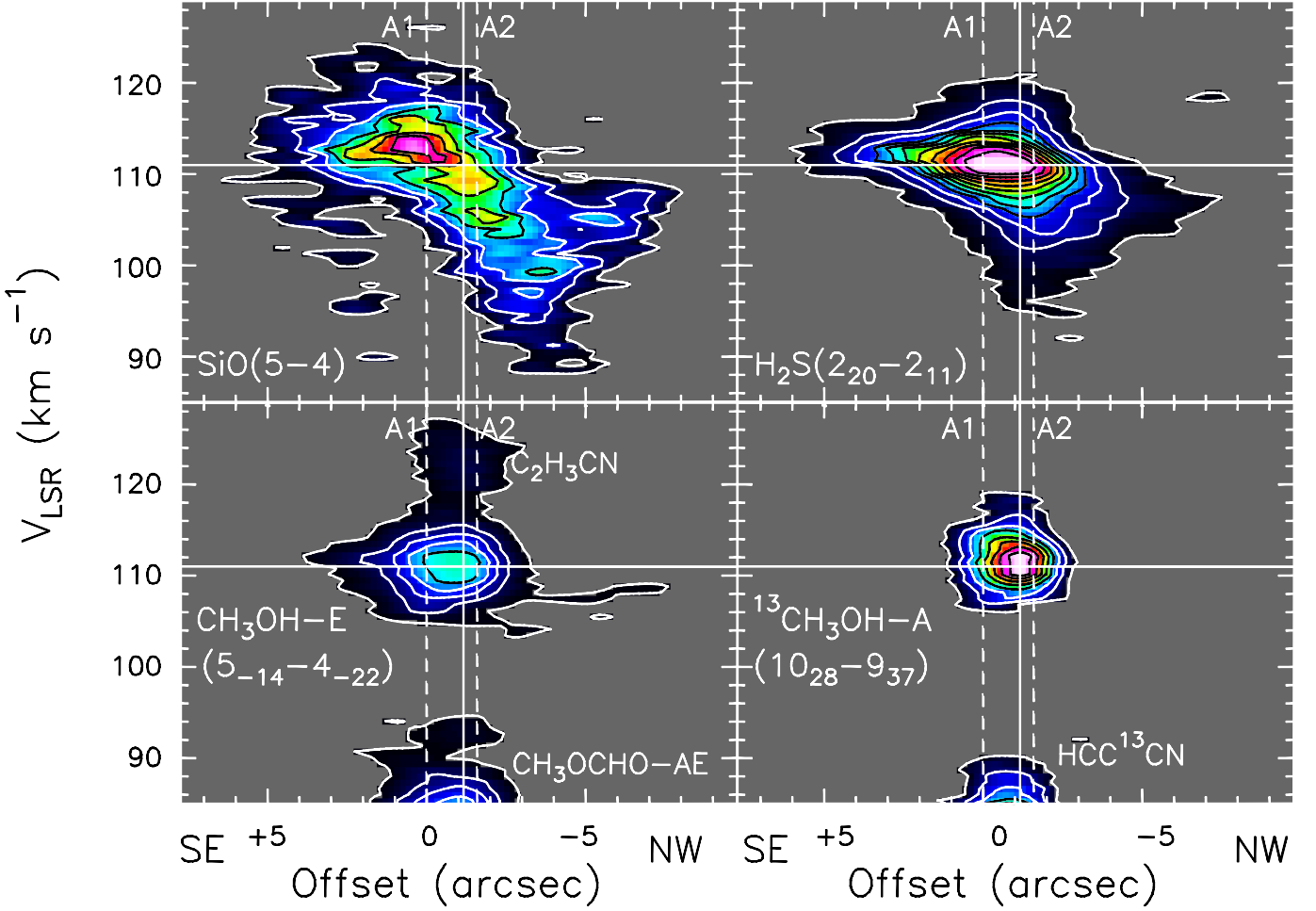}
\caption{Position-velocity cut of SiO(5--4), H$_2$S(2$_{2,0}$--2$_{1,1}$),
CH$_3$OH-E(5$_{\rm -1,4}$--4$_{\rm -22}$), and 
$^{13}$CH$_3$OH--A(10$_{\rm 2,8}$--9$_{\rm 3,7}$) along the 
outflow A (PA = --40$\degr$). 
The position offsets are measured from the A1 position, positive towards 
northwest.
The rms 1$\sigma$ is 20 mJy beam$^{-1}$.
Contour levels range from 3$\sigma$ by steps of 3$\sigma$ for
SiO and $^{13}$CH$_3$OH--A, and from 3$\sigma$ by steps of 10$\sigma$
for H$_2$S and CH$_3$OH-E.
Dashed vertical lines mark the position of the A1 and A2 YSOs,
while the solid vertical line is for 
the average peak position of the typical hot-core tracers (see text). 
Solid horizontal line shows the systemic velocity (+111.0 km s$^{-1}$).}
\label{pvall}
\end{figure*}

\begin{figure*}
\centering
\includegraphics[angle=-90,width=16cm]{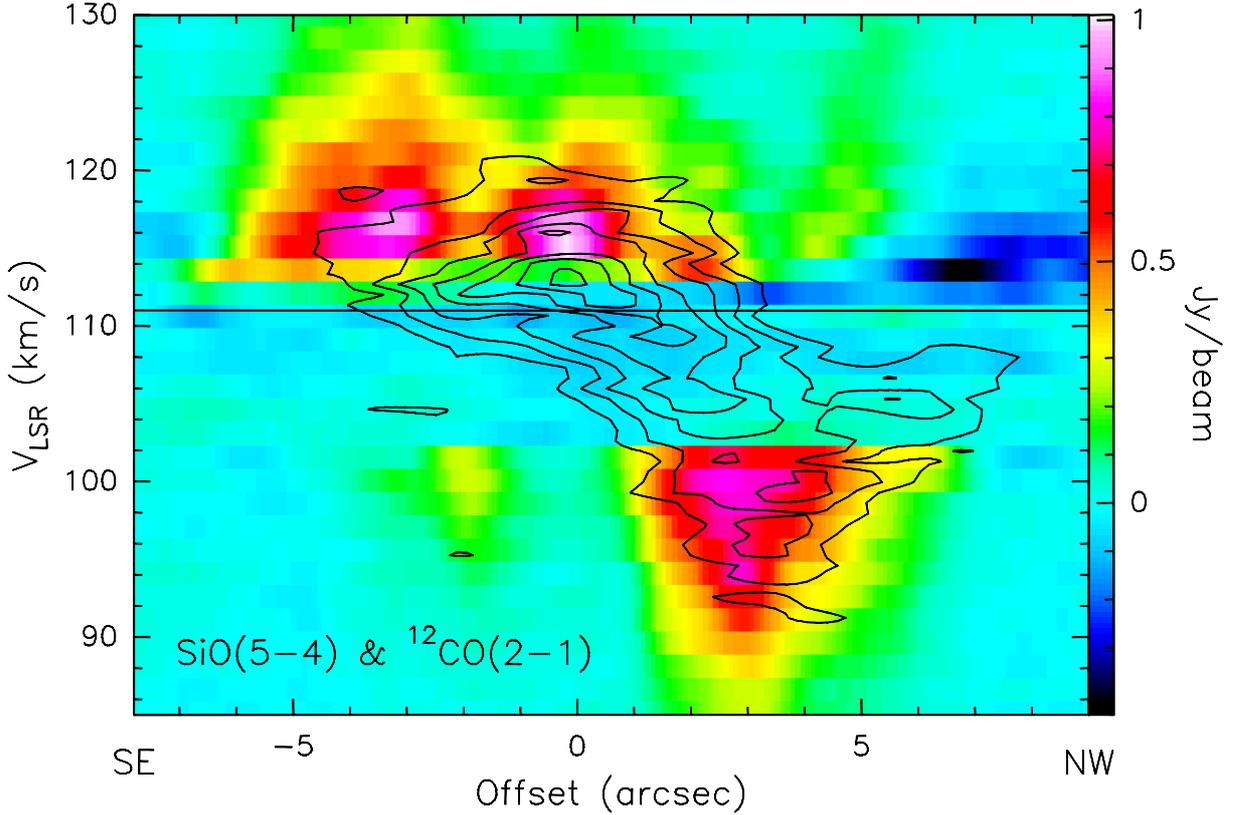}
\caption{Comparison between the position-velocity cut of SiO(5--4), black contours,
and that of CO(2--1) in colours, derived along the 
outflow A (PA = --40$\degr$) using the CO dataset
presented by Beltr\'an et al. (2011).   
The solid horizontal line shows the systemic velocity (+111.0 km s$^{-1}$).}
\label{pscosio}
\end{figure*}

\begin{figure}
\centering
\includegraphics[angle=-90,width=8cm]{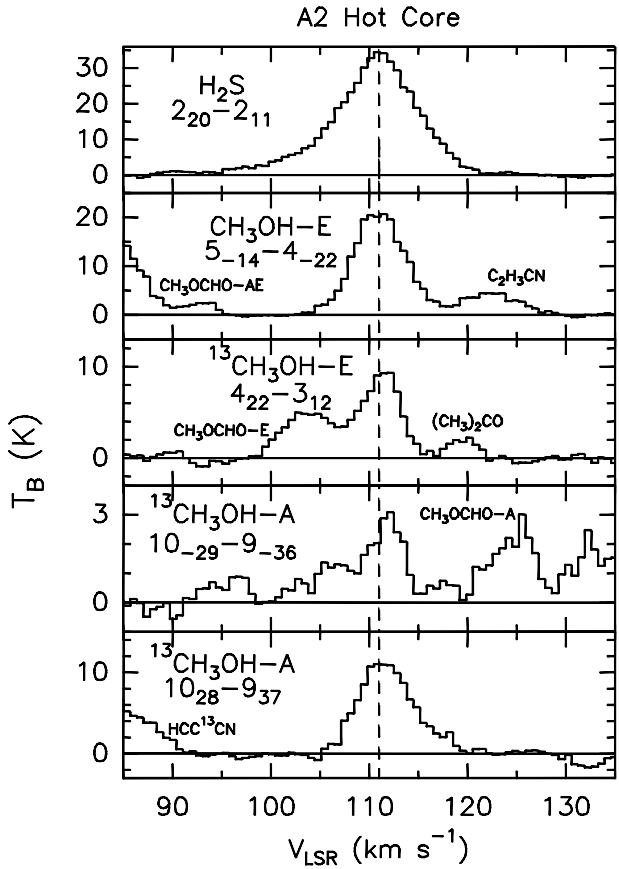}
\caption{SiO, H$_2$S, CH$_3$OH, and $^{13}$CH$_3$OH line profiles
(in brightness temperature, $T_{\rm B}$, scale) observed towards the A2 hot core
with the SMA.
The dashed lines denote the systemic velocity (+111 km s$^{-1}$).
Several high-excitation lines due to typical tracers
of hot-core chemistry (methyl formate,
acetone, vinyl cyanide, cyanoacetilene) have also been detected (see Table 4).}
\label{spectrasma}
\end{figure}

\begin{figure}
\includegraphics[angle=-90,width=8cm]{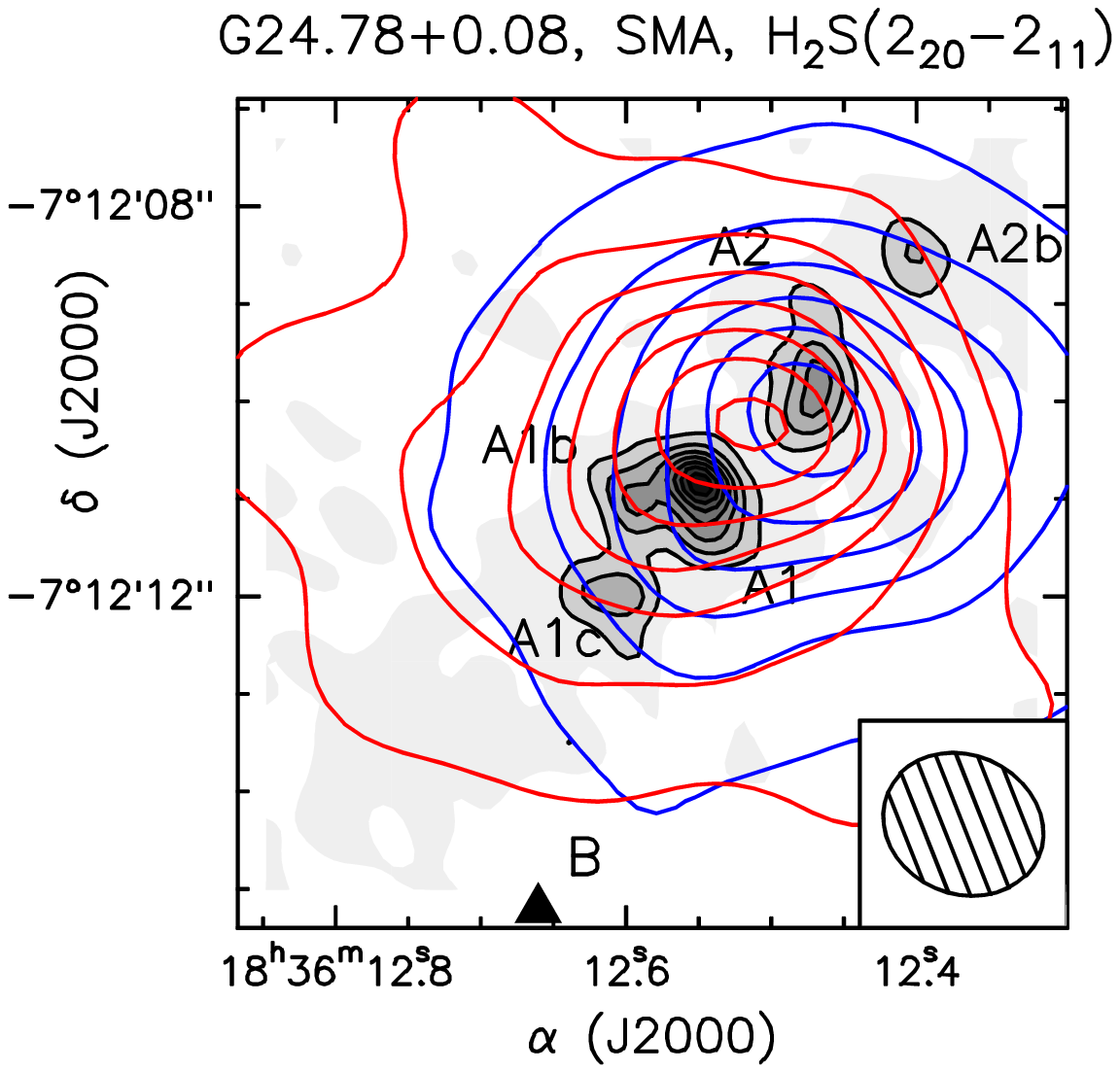}
\caption{Contour map of blue- and red-shifted H$_2$S(2$_{2,0}$--2$_{1,1}$)
SMA emission
superimposed on the 1.3 mm continuum emission as observed at high angular
resolution by Beltr\'an et al. (2011) using a very extended SMA configuration
($0\farcs55\times0\farcs44$). The sources of the G24.78+0.08 cluster
are labelled following Furuya et al. (2002) and Beltr\'an et al. (2004, 2011).
The H$_2$S emission was averaged over the velocity intervals (+94,+111) km s$^{-1}$
and (+111,+124) km s$^{-1}$ for the blue- and red-shifted emission, respectively.
The rms 1$\sigma$ is 18 mJy beam$^{-1}$ km s$^{-1}$.
Contour levels range from 5$\sigma$ by steps of 10$\sigma$.
The filled ellipse in the bottom-right corner shows the
synthesised beam (HPBW): $1\farcs7\times1\farcs4$ (PA =
67$\degr$).}
\label{h2s}
\end{figure}

\section{Physical conditions as traced by SiO}

\subsection{SiO(1--0) absorption}

From the SiO(1--0) absorption feature one can obtain an estimate of the
excitation temperature of this transition. We assume that the absorption
is due to the blue lobe of the outflow lying between the observer and the
hypercompact \HII\ region. Since the lobe is resolved in our maps, while
the \HII\ region is much smaller than the synthesised beam of the SiO images,
the SiO(1--0) brightness temperature measured in the synthesised beam
is a mixture of absorption (towards the HII region) and emission (from
the rest of the beam):
\begin{equation}
 \Ob \Tb = \left(\Ob-\Os\right) \Te + \Os \Ta,
\end{equation}
where \Te\ and \Ta\ are the brightness temperatures of the SiO(1--0) line in
emission and absorption, while \Os\ and \Ob\ are the solid angle subtended
by the \HII\ region and the synthesised beam of the SiO(1--0) map.
Moreover, \Te\ and \Ta\ are given by
\begin{eqnarray}
\Te & = & \Tex \left(1-\e^{-\taue}\right) \label{ete} \\
\Ta & = & \left(\Tex-\Thii\right) \left(1-\e^{-\taua}\right),
\end{eqnarray}
with \Tex\ excitation temperature of the $J$=1--0 transition and
\Thii\ brightness temperature of the continuum emission from the hypercompact
\HII\ region. Here, we have assumed that \Tex\ is constant along the line of
sight and across the region covered by the synthesised beam.
After some algebra, one obtains the expression
\begin{eqnarray}
\Tb & = & \Tex \left(1-\e^{-\taue}\right) \left[ 1-\frac{\Os}{\Ob}
	   \frac{\e^{-\taua}-\e^{-\taue}}{1-\e^{-\taue}}
	   \right] \nonumber \\
	  &   &  - \frac{\Os}{\Ob} \Thii \left(1-\e^{-\taua}\right).
	    \label{etb}
\end{eqnarray}
We note that the \Ob\ is relatively small compared to the size of the
outflow lobe, and hence the opacity should not depend significantly
on the line of sight. Also, for obvious reasons, the \HII\ region either
lies
behind the lobe or is enshrouded by it. Therefore, $\taua\le\taue$ and
$0\le\frac{\e^{-\taua}-\e^{-\taue}}{1-\e^{-\taue}}<1$.
Since $\frac{\Os}{\Ob}\ll1$, Eq.~(\ref{etb}) can be written as
\begin{equation}
\Tb \simeq \Tex \left(1-\e^{-\taue}\right) -
\frac{\Os}{\Ob} \Thii \left(1-\e^{-\taua}\right). \label{etba}
\end{equation}
Most likely, the absorption and emission lines of sight are crossing similar
amounts of gas, and we hence
make the additional assumption that $\taua\simeq\taue$. By replacing
Eq.~(\ref{ete}) into Eq.~(\ref{etba}),
one can finally obtain an expression for \Tex:
\begin{equation}
\Tex = \frac{\Os}{\Ob}\Thii \frac{\Te}{\Te-\Tb}.
\end{equation}

From our observations we measure absorption with $\Tb\simeq-6$~K in a beam
of $\Ob=4.24$~arcsec$^2$, while from the data of Beltr\'an et al. (2007) one
obtains $\Os\simeq0.0554$~arcsec$^2$ and a source-averaged brightness
temperature of the \HII\ region at 7~mm $\Thii\simeq960$~K. An estimate of \Te\
can be obtained from the SiO(1--0) spectra one beam away from the absorption
peak: $\Te \simeq$ 3--6~K. With these numbers we obtain $\Tex \simeq 4$--8~K.
The brightness temperature of the continuum emission measured at 7~mm
towards the \HII\ region in our images is $\sim$12~K, i.e. comparable to
the value ($\frac{\Os}{\Ob}\Thii\simeq12$~K) estimated applying beam dilution
to the high-resolution measurement of Beltr\'an et al. (2007): this proves
that basically the entire continuum emission in our beam is due to free-free
emission from the ionised gas, while dust emission is negligible.

\subsection{LVG analysis}

We ran the RADEX non-LTE model (van der Tak et al. 2007) with the
rate coefficients for collisions with H$_2$ reported by
Turner et al. (1992) using the escape probability method for a 
plane-parallel geometry to fit the observed SiO line ratios and brightness.
We explored \H2\
densities from $10^2$ to $10^8$ cm$^{-3}$, kinetic temperatures $T_{\rm kin}$ from
50 to 500 K, and
an LVG optical depth parameter $n$(SiO)/ $(dV/dz)$ = \NdV\ ranging
from $10^{12}$ to $10^{17}$ \cmsq\ (\kms)$^{-1}$, i.e. from the fully
optically thin to the optically thick regime.
We used an FWHM linewidth of 10 km s$^{-1}$, as suggested by
the SiO spectra (see Fig. 4).

As a first step, we modelled the SiO emission observed with the IRAM
30-m telescope, using the SiO(2--1) and (5--4) transitions, which are sensitive to  
similar excitation conditions (see Table 1) as the lines observed with the interferometers
(discussed below).
The low-J SiO lines, as already learned from previous studies
of low-mass protostellar systems (e.g. Cabrit et al. 2007 and references therein),
are not very sensitive to kinetic temperature.
Figure 13 thus reports the solutions for the observed SiO(5--4)/SiO(2--1)
intensity ratio at the typical red- (+116 km s$^{-1}$) and blue-shifted (+106 km s$^{-1}$)
emission in the $n_{\rm H_2}$--$N$(SiO) plane and for two extreme temperatures: 50
and 500 K.
The observed intensity ratio was corrected for beam dilution  
assuming an emitting source of 7$\arcsec$, which is
representative of the sizes of the outflows imaged with the VLA and SMA.
To show how much the results depend on this assumption on emitting
size, we also report the solutions for a smaller size (i.e. 3$\arcsec$), which
is the typical size of the brightest SiO clumps in G24.
The dashed contours take into account the uncertainties associated with the intensity ratio.
The 30-m spectra do not provide any constraint on the kinetic temperature and
column density, while densities of below 10$^{5}$ cm$^{-3}$
can be inferred.

On the other hand, we can obtain tight constraints on volume and column
densities by analysing
the SiO emission observed with the VLA and SMA arrays, after degrading
the SMA map to the angular resolution of the VLA image.
Figures 14 and 15 report the
solutions for the observed SiO(1--0) and SiO(5--4) brightness temperatures
at the same red- (+116 km s$^{-1}$) and blue-shifted (+106 km s$^{-1}$)
velocities investigated using the 30-m spectra.
We modelled the emission observed towards three positions: A1 (Fig. 14), as
well as where the J=1--0 emission is only weakly 
affected by absorption, i.e. at the position of the red- and blue-shifted
SiO(1--0) emission peaks (Fig. 15).
As reported in Table 3, the positions of
the red-shifted peak as traced by SiO(1--0) and (5--4), hence once
considered the angular resolutions, agree well.
We report the solutions found for the same
kinetic temperatures used above to model the 30-m emission (50 and 500 K).
We plot in black in Fig. 14 the contours corresponding to the
the excitation temperature of the SiO(1--0) line estimated
in Sect. 4.1 from the SiO(1--0) absorption feature observed towards A1.

In practice, the present LVG plots indicate 
for the three positions (i) volume densities between 10$^3$ and 10$^5$ cm$^{-3}$,
and (ii) well-constrained SiO column density,  
in the 0.5--1 10$^{15}$ cm$^{-2}$ range. 
These solutions are consistent with what was found using the
30-m IRAM SiO(5--4) and (2--1) spectra.
Nisini et al. (2007) performed an LVG analysis of SiO emission towards   
the nearby prototypical Class 0 low-mass objects L1148 and L1157,
using single-dish (IRAM, JCMT) data on angular scales of 10$\arcsec$. 
In that case as well, no tight constraints have been obtained for the kinetic temperature,
whereas the volume densities were found to be between 10$^5$ and 10$^6$ cm$^{-3}$. 
Moreover, Gusdorf et al. (2008a) modelled the SiO emission from L1157 
using a C-shock code, which led to pre-shocked densities of
10$^4$--10$^5$ cm$^{-3}$. Thus, although completely different spatial scales are involved, 
the volume densities here inferred 
for G24 (10$^3$--10$^5$ cm$^{-3}$) appear to be consistent with those
derived for L1448 and L1157. 
 
Assuming that the physical conditions reported above are representative
of the whole SiO outflow (including where SiO(1--0) emission is affected by absorption),
using a volume density of 10$^{4}$ cm$^{-3}$, the inferred size of the lobes (7$\arcsec$), 
and the G24 distance (7.7 kpc), the corresponding outflow mass is
$\sim$ 40 $M_{\rm \sun}$. 
This implies, according to Beuther et al. (2002),
Zhang et al. (2001, 2005), and
L\'opez-Sepulcre et al. (2009, see their Fig. 5), a luminosity for the driving source of the SiO outflow
of a few 10$^{4}$ $L_{\rm \sun}$, corresponding to a late O-type ZAMS YSO (e.g. Panagia 1973).         
If we conservatively assume $n_{\rm H_2}$ = 10$^{3}$ cm$^{-3}$, the driving
source is still a massive YSO, corresponding to a B2 spectral type. 
However, for consistency purposes, we cross-checked the outflow masses using  
the estimates of SiO column density and assuming a given SiO abundance relative to H$_2$.  
The latter has a large uncertainty since SiO can be greatly
enhanced in shocks. Assuming typical abundances of 10$^{-8}$--10$^{-7}$
(i.e. 4--5 orders of magnitude greater than that in dark clouds,
10$^{-12}$, Ziurys et al. 1989),  
we inferred volume densities between 10$^4$ and 10$^5$ cm$^{-3}$, and thus 
again masses ($\sim$ 40--400 $M_{\rm \sun}$) in agreement with an outflow 
driven by a late O-type ZAMS YSO. 

\begin{figure*}
\centering
\includegraphics[angle=-90,width=18cm]{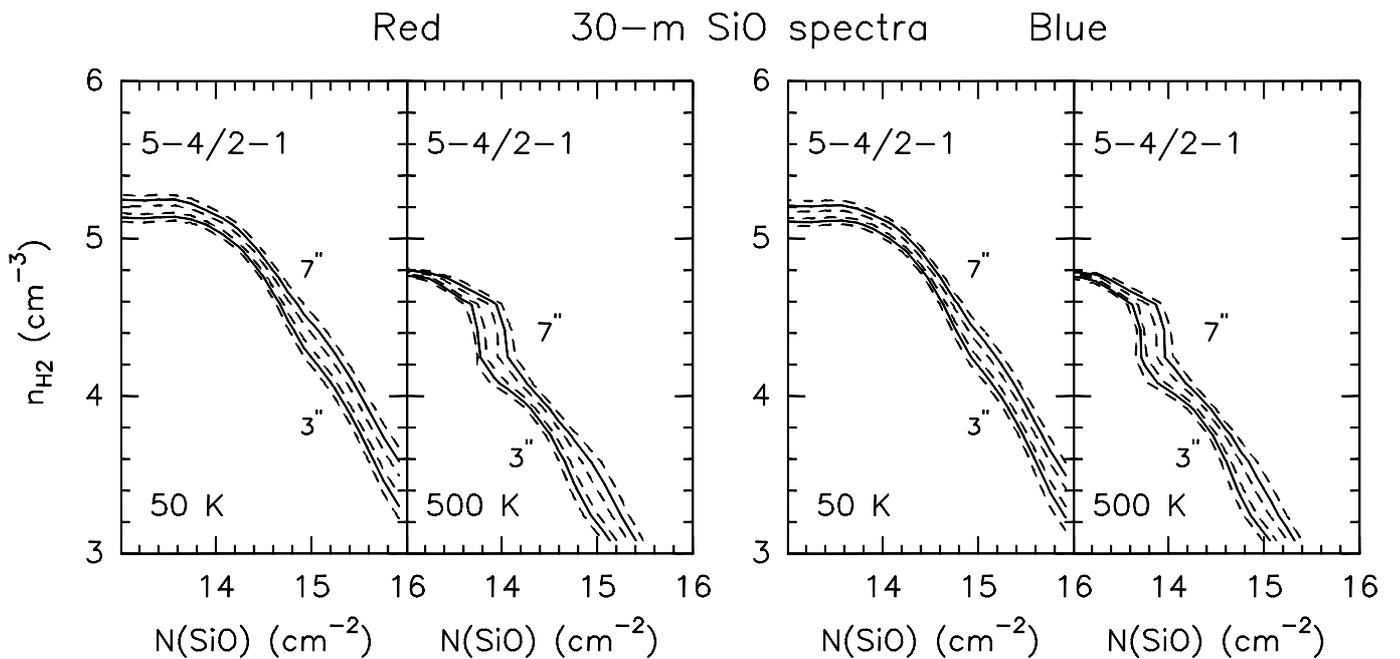}
\caption{Analysis of the SiO
red-shifted line emission in the G24.78+0.08 outflow A 
observed with the IRAM 30-m antenna.
The solutions for the observed SiO(5--4)/SiO(2--1) intensity ratio
are shown in the $n_{\rm H_2}$--$N$(SiO) plot for non-LTE (RADEX) 
plane-parallel models at the labeled kinetic temperatures.
Solid contours are for the measured ratios
at the typical red- (+116 km s$^{-1}$) and blue-shifted (+106 km s$^{-1}$)
velocities, after correction for a beam dilution derived  
assuming a source size of 3$\arcsec$ and 7$\arcsec$ (see text).
The SiO(5--4)/SiO(2--1) ratios are 0.21 (3$\arcsec$; +116 km s$^{-1}$),
0.19 (3$\arcsec$; +106 km s$^{-1}$), 0.27 (7$\arcsec$; +116 km s$^{-1}$),
and 0.30 (7$\arcsec$; +106 km s$^{-1}$).
Dashed contours take into account the uncertainties.}
\label{radex30m}
\end{figure*}

\begin{figure*}
\centering
\includegraphics[angle=-90,width=18cm]{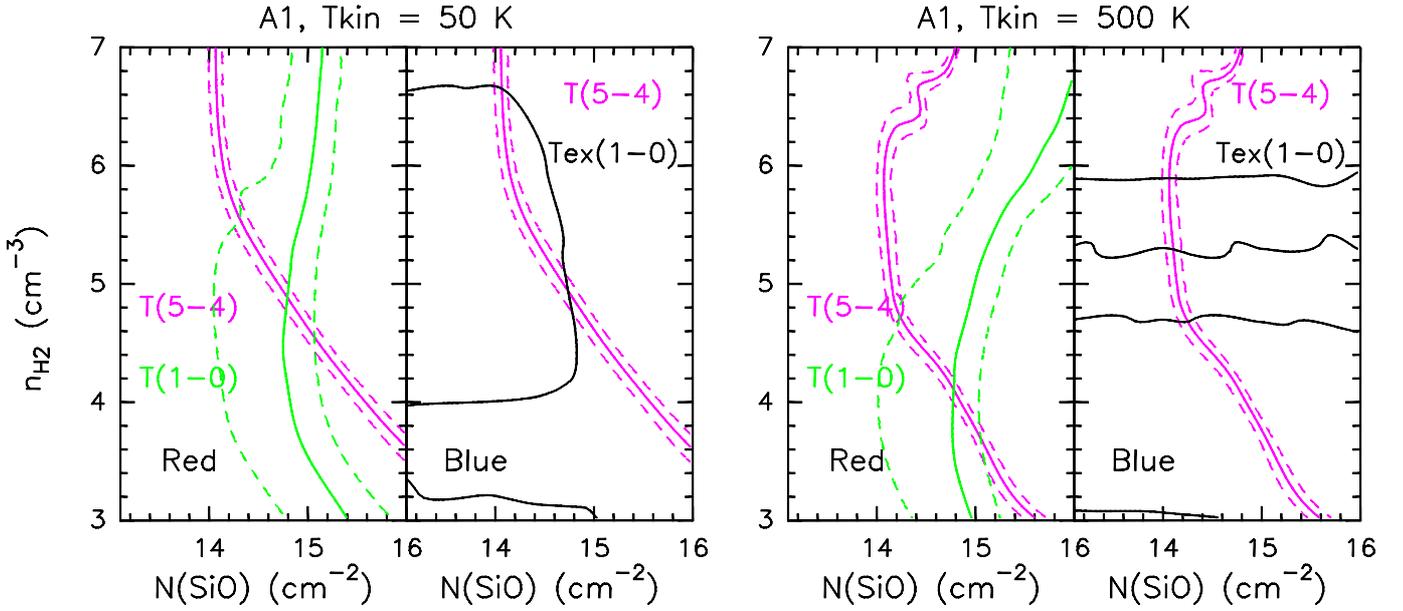}
\caption{Analysis of the SiO line blue-
and red-shifted emissions towards A1 as
observed with VLA and SMA (convolved to the angular resolution of the VLA
image).
The solutions for the observed SiO(1--0) and SiO(5--4) 
brightness temperatures (solid
lines) 
at the typical red- (+116 km s$^{-1}$) and blue-shifted (+106 km s$^{-1}$) 
velocities 
are shown in the $n_{\rm H_2}$--$N$(SiO) plane for non-LTE (RADEX) 
plane-parallel models at the labelled kinetic temperatures.
Dashed lines are for the uncertainties.
The SiO(5--4) brightness temperatures are 4.4 K (+116 km s$^{-1}$) and
4.3 K (+106 km s$^{-1}$), while the SiO(1--0) brightness temperature is  
4.0 K (+116 km s$^{-1}$).
Given the blue-shifted absorption observed towards A1, 
we plot in black the solutions corresponding to
the derived SiO(1--0) excitation temperature ($\simeq$ 6 K, see Sect. 4.1).}
\label{radexa1}
\end{figure*}

\begin{figure*}
\centering
\includegraphics[angle=-90,width=18cm]{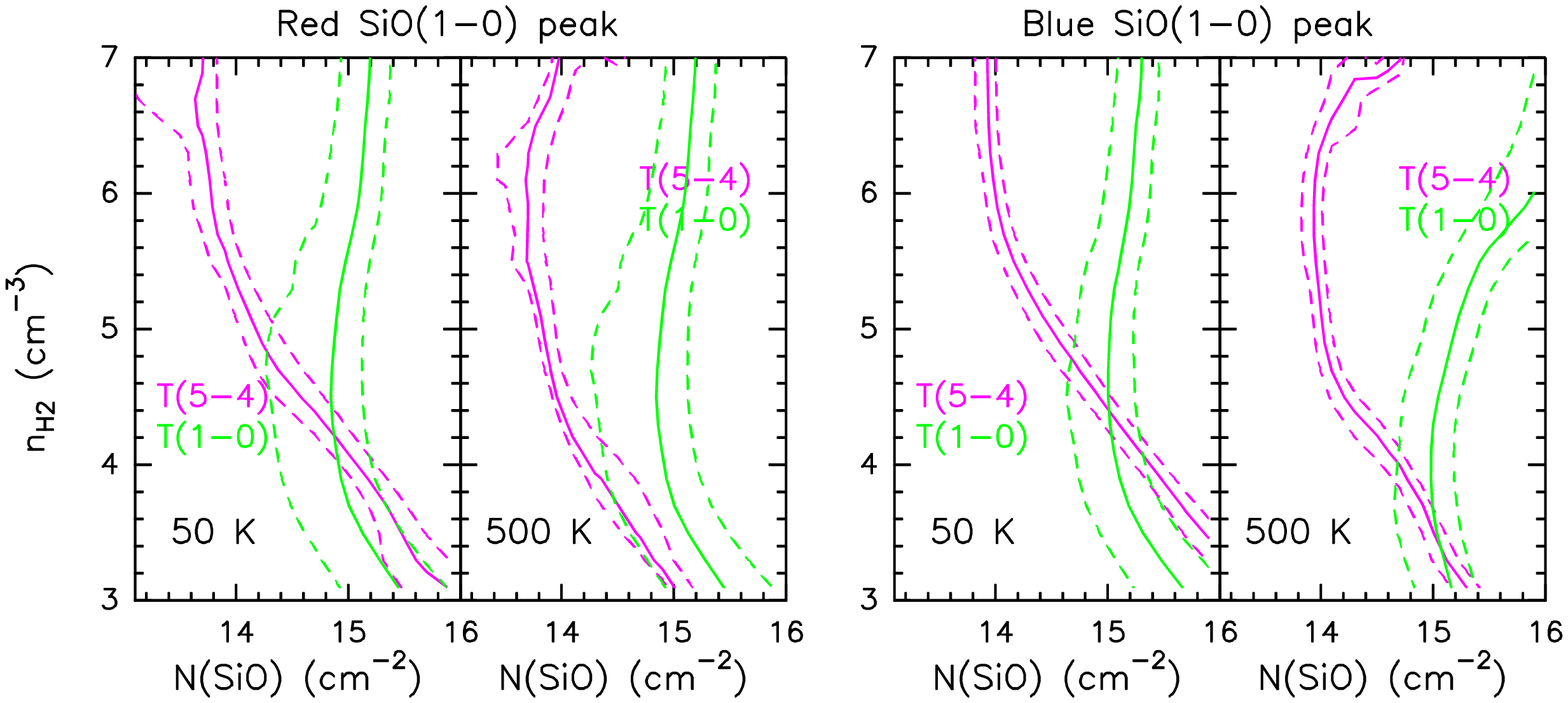}
\caption{Analysis of the SiO line blue-
and red-shifted emissions towards the corresponding  
emission peaks of the SiO(1--0) map (see text), as  
observed with VLA and SMA (convolved to the angular resolution of the VLA
image).
The solutions for the observed SiO(1--0) and SiO(5--4) brightness
temperatures 
at the typical red- (+116 km s$^{-1}$) and blue-shifted (+106 km s$^{-1}$)
velocities
are shown in the $n_{\rm H_2}$--$N$(SiO) plane for non-LTE (RADEX) 
plane-parallel models at the labelled kinetic temperatures.
The SiO(1--0) brightness temperatures are 4.7 K (+116 km s$^{-1}$) and
6.1 K (+106 km s$^{-1}$), while the SiO(5--4) brightness temperatures are
1.1 K (+116 km s$^{-1}$) and 2.9 K (+106 km s$^{-1}$).
Dashed lines are for the uncertainties.}
\label{radexredblue}
\end{figure*}

\section{Conclusions}

We conducted a multiline SiO survey towards the G24.78+0.08 region,
which is an excellent laboratory to study the process of
high-mass star formation, because it is associated with YSOs in different
evolutionary stages. After preliminary IRAM 30-m single-dish runs, 
we obtained high angular resolution images using the VLA and SMA interferometers. 
The main results can be summarised as follows: 

\begin{enumerate}

\item
High-velocity SiO emission (up to 25 km s$^{-1}$ w.r.t. the systemic
velocity, +111 km s$^{-1}$) reveals two collimated outflows driven by the A2 and C millimeter continuum
massive cores. On the other hand, no SiO emission has been detected  
towards more evolved young stellar objects associated 
with an UC\HII\ region (core B) or driven by the hypercompact (core A1) \HII\ regions.
Moreover core D shows no SiO emission, confirming its quiescent nature,
without any signature of star formation. 

\item
The LVG analysis of the SiO emission reveals high-density gas
(10$^{3}$--10$^{5}$ cm$^{-3}$), with clearly constrained SiO column
densities ($\sim$ 10$^{15}$ cm$^{-2}$).
The average velocity of the SiO emission increases
as a function of distance from the driving source, which suggests gas acceleration.
Although the angular resolution is not high
enough to demonstrate the occurrence of SiO jets, if we assume the standard approach that SiO
is tracing shocks inside jets, it is reasonable to associate  
the observed collimated SiO structures with jet activity.

\item
The driving source of the A2 outflow 
(i) has an estimated luminosity of $\ge$ 10$^{4}$ $L_{\rm \sun}$, which is typical of a late O-type star,
and (ii) is located at the centre of a hot molecular core
(traced in the present data set by 
emission from methyl formate, vinyl cyanide, 
cyanoacetilene, and acetone)  
that rotates on a plane perpendicular to the outflow main axis.

\item
To our knowledge, we obtained one of the first interferometric images 
of an SiO jet-like outflow from young $\ge$ 10$^{4}$ $L_{\rm \sun}$ stars. 
High spatial resolution maps of SiO high-velocity emission 
driven from young $\ge$ 10$^{4}$ $L_{\rm \sun}$ stars have
have been so far obtained towards IRAS18264--1152, 18566+0408, 20126+4104, 
and 23151+5912 (Cesaroni et al. 1999; Qiu et al. 2007; Zhang et al. 2007).  
IRAS20126+4104 is probably  
the best example of a jet from a massive YSO so far observed; it is 
traced by SiO, H$_2$, and  [FeII] emission (Caratti o Garatti et al. 2008).
To conclude, the present SiO observations support the theory 
that O-type stars form according to a core accretion model, i.e. via a scaled-up picture typical
of Sun-like star formation, where
jets, well-traced by SiO emission, create outflows of accumulated and accelerated 
ambient gas that in turn is well traced by CO.

\end{enumerate}

\begin{acknowledgements}
We wish to thank the IRAM staff for the 30-m observations performed
in service mode. We are grateful to S. Cabrit for useful discussion and suggestions.
We also thank the anonymous referee for comments and suggestions, which  
improved the work.
\end{acknowledgements}



\end{document}